\documentclass[letterpaper, 10 pt, journal, twoside]{IEEEtran}
\usepackage{amsmath,amsfonts, amssymb}
\usepackage{algorithmic}
\usepackage{algorithm}
\usepackage{array}
\usepackage[caption=false,font=footnotesize]{subfig}
\usepackage{titlesec}
\usepackage{textcomp}
\usepackage{stfloats}
\usepackage{url}
\usepackage{verbatim}
\usepackage{graphicx}
\usepackage{cite}
\usepackage{ragged2e}
\usepackage{amsthm}
\usepackage{makecell} 
\usepackage{mathrsfs}
\usepackage{mathabx} 
\usepackage{xcolor}
\usepackage{bbm}
\usepackage{multirow}
\usepackage[hidelinks]{hyperref}
\usepackage{cleveref}
\usepackage{booktabs}
\usepackage[T1]{fontenc}  
\usepackage[
  paper=letterpaper,
  left=0.68in,
  right=0.68in,
  top=0.833in,
  bottom=0.68in 
]{geometry}

\makeatletter
\renewenvironment{proof}[1][\proofname]{%
  \par\vspace{-0.05em}
  \pushQED{\qed}%
  \normalfont
  \topsep0pt \partopsep0pt 
  \trivlist
  \item[\hskip\labelsep\itshape\bfseries #1\@addpunct{.}]%
}{%
  \popQED\endtrivlist\@endpefalse
  \vspace{-0.1em}
}
\makeatother

\titlespacing*{\subsection}
  {0pt}{0.4\baselineskip}{0.2\baselineskip}

 \titlespacing*{\section}
   {0pt}{0.9\baselineskip}{0.3\baselineskip}

\allowdisplaybreaks

\newtheoremstyle{tightassumption}%
  {1pt}   
  {1pt}   
  {}      
  {}      
  {\bfseries} 
  {.}     
  {0.5em} 
  {}      
\theoremstyle{tightassumption}
\newtheorem{assumption}{Assumption} 
\newtheoremstyle{tighttheorem}%
  {1pt}   
  {1pt}   
  {}      
  {}      
  {\bfseries} 
  {.}     
  {0.5em} 
  {}      
\theoremstyle{tighttheorem}
\newtheorem{theorem}{Theorem}

\newtheorem{lemma}{Lemma} 
\newtheorem{proposition}{Proposition} 
\newtheoremstyle{tightremark}%
  {1pt}   
  {1pt}   
  {}      
  {}      
  {\bfseries} 
  {.}     
  {0.5em} 
  {}      
\theoremstyle{tightremark}
\newtheorem{remark}{Remark}

\newcommand{\spanop}{\text{span}}

\newcommand{\vect}[1]{\boldsymbol{#1}}

\def\bs{\boldsymbol}
\def\mf{\mathbf}
\def\mb{\mathbb}
\def\mc{\mathcal}
\def\beq{\begin{equation*}}
\def\eeq{\end{equation*}}
\def\bql{\begin{equation}}
\def\eql{\end{equation}}
\def\bqn{\begin{eqnarray*}}
\def\eqn{\end{eqnarray*}}
\def\bnl{\begin{eqnarray}}
\def\enl{\end{eqnarray}}
\def\bna{\bql\begin{array}{rcl}}
\def\ena{\end{array}\eql}
\def\bnn{\beq\begin{array}{rcl}}
\def\enn{\end{array}\eeq}
\def\bma{\begin{bmatrix}}
\def\ema{\end{bmatrix}}
\def\bmx{\begin{matrix}}
\def\emx{\end{matrix}}
\def\ben{\begin{enumerate}}
\def\een{\end{enumerate}}
\def\bit{\begin{itemize}}
\def\eit{\end{itemize}}
\def\bei{\begin{itemize}}
\def\eei{\end{itemize}}
\def\bet{\begin{tabular}}
\def\eet{\end{tabular}}

\newcommand{\allcaps}[1]{\uppercase\expandafter{#1}}

\providecommand{\norm}[1]{\left\|#1\right\|}

\def\SE{\operatorname{SE}}
\def\SO{\operatorname{SO}}

\begin{document}

\setlength{\abovedisplayskip}{0.6pt}
\setlength{\belowdisplayskip}{0.6pt}
\setlength{\textfloatsep}{0.6pt}   
\setlength{\intextsep}{0.6pt}
\setlength{\floatsep}{0.6pt} 

\title{Real-Time Linear MPC for Quadrotors on SE(3): An Analytical Koopman-based Realization}

\author{ Santosh Rajkumar$^{1}$, Chengyu Yang$^{2}$, Yuliang Gu$^2$, Sheng Cheng$^{2}$, Naira Hovakimyan$^{2}$, and Debdipta Goswami$^{1}$
\thanks{This work is supported by the OSU Presidential Research Excellence (PRE) Accelerator grant, NSF-DMS Math-DT (2529302), NASA cooperative agreement (80NSSC22M0070), AFOSR (FA9550-21-1-0411), NSF-AoF Robust Intelligence (2133656), and NSF SLES (2331878)}
\thanks{$^{1}$Authors are with the Department of Mechanical and Aerospace Engineering, The Ohio State University, Columbus,
OH, USA.
Email: {\tt {\{rajkumar.36, goswami.78\}@osu.edu}}. }
\thanks{
$^{2}$Authors are with the Department of Mechanical Science and Engineering, University of Illinois, Urbana-Champaign, IL, USA. Email:{ \tt \{cy45, yuliang3, chengs, nhovakim\}@illinois.edu }}
\thanks{Code and demo(s) available at \href{https://github.com/santoshrajkumar/kq-lmpc-quadrotor}{https://github.com/santoshrajkumar/kq-lmpc-quadrotor}. }
}



\maketitle
\begin{abstract}
This letter presents an analytical linear parameter-varying (LPV) representation of quadrotor dynamics utilizing Koopman theory, facilitating computationally efficient linear model predictive control (LMPC) for real-time trajectory tracking. By leveraging carefully designed Koopman observables, the proposed approach enables a compact lifted-space evolution that mitigates the curse of dimensionality while preserving the nonlinear characteristics of the system. Although model predictive control (MPC) is a powerful strategy for quadrotor control, it faces a trade-off between the high computational cost of nonlinear MPC (NMPC) and the reduced accuracy of LMPC. To address this gap, we introduce KQ-LMPC (Koopman Quasilinear LPV MPC), which leverages the Koopman-lifted LPV formulation to enforce constraints, ensure lower computational burden and real-time feasibility, and deliver tracking performance comparable to NMPC. Experimental validation confirms the effectiveness of the framework in reasonably agile flight. To the best of our knowledge, this is the first experimentally validated LMPC for quadrotors that employs analytically derived Koopman observables without requiring training data.

\end{abstract}


\vspace{-0.6em}
\section{Introduction}
\IEEEPARstart{Q}{uadrotors}, recognized as a type of mobile robot and unmanned aerial vehicle (UAV), are increasingly used in industry due to their streamlined design and the ability to execute complex maneuvers \cite{mohsan2023unmanned}. Quadrotor control has received significant attention in research due to the complex nonlinear dynamics of the vehicle and the inherently underactuated nature \cite{emran2018review}. Desirable features for quadrotor control include: (i) accurate trajectory tracking with future reference anticipation, (ii) handling system constraints, (iii) robustness to disturbances and unmodeled dynamics, and (iv) real-time feasibility given their fast dynamics \cite{greeff2018flatness}. However, it remains challenging to find all these features in a single method \cite{andrien2024model}.


In this context, Model Predictive Control (MPC) has emerged as a promising approach for quadrotor control \cite{mayne2000constrained}, with NMPC effectively addressing nonlinear dynamics and enabling agile flight maneuvers \cite{wang2021efficient, pereira2021nonlinear}. Comparative studies show that NMPC achieves superior tracking accuracy compared to differential flatness-based methods, but at a substantially higher computational cost \cite{sun2022comparative}. This computational burden arises from solving a non-convex nonlinear program (NLP) at each step \cite{primbs1997mpc}, making real-time feasibility especially challenging on embedded platforms with limited resources and for safety-critical applications requiring high-frequency control updates. While cascaded control architectures have been proposed to mitigate this challenge \cite{schlagenhauf2020cascaded, andrien2024model}, they complicate constraint handling and can degrade agile performance due to task separation \cite{pereira2021nonlinear}. Another strategy is single-iteration Sequential Quadratic Programming (SQP), which reduces computation but may yield approximate solutions that can violate state or input constraints, posing risks in safety-critical scenarios \cite{wang2009fast}. Despite NMPC’s demonstrated advantages, ensuring reliable real-time implementation remains insufficiently addressed in the literature.



On the other hand, LMPC replaces the non-convex NLP with a tractable quadratic program (QP), ensuring predictable solving times and guaranteed convergence to the global optimum within the convex formulation \cite{gros2020linear}. This makes LMPC computationally efficient and well-suited for real-time deployment. However, this efficiency comes at the cost of reduced accuracy and potential instability due to its dependence on linearized dynamics \cite{bangura2014real}. Local linearization techniques \cite{alexis2011model, islam2017dynamics} perform adequately near hover but deteriorate with larger deviations. To address this, LMPC schemes combining differential flatness-based feedback and feedforward linearization in cascaded structures have been proposed \cite{greeff2018flatness, martins2022inner, nwafor2024optimal}. However, these methods typically require smooth reference trajectories and struggle with explicit constraint handling. Control barrier function-based quadratic programs (CBF-QPs) \cite{khan2025trajectory} improve constraint enforcement and enable smooth reference generation, but their reliance on Euler angles introduces gimbal lock, limiting suitability for agile maneuvers \cite{wu}.

Thus, designing MPC schemes for quadrotor control inherently involves a trade-off between the high computational cost and real-time reliability risks associated with the superior accuracy of NMPC, and the reduced accuracy of LMPC chosen for the sake of real-time computational reliability. Bridging this gap by developing a controller that achieves tracking performance reasonably comparable to NMPC while maintaining the computational efficiency and reliability of LMPC remains a central challenge in advancing high-performance, safety-critical agile flight. Koopman operator theory offers a promising path forward by providing (almost) globally linear representations of nonlinear dynamics in a lifted space of observables \cite{ budivsic2012applied}, enabling efficient LMPC implementation for quadrotor control.

Koopman operator theory provides a linear representation of nonlinear dynamics in terms of the evolution of a set of functions, termed \emph{observables}, in a Banach space \cite{koopman1931hamiltonian}. In practice, one must specify a dictionary of observables whose span defines (or approximates) a Koopman-invariant subspace, referred to as the \emph{lifted space}. However, identifying the relevant observables that define a Koopman-invariant subspace is not always straightforward \cite{brunton2016koopman}.  State-of-the-art Koopman-based approaches for quadrotor control are typically based on data-driven methods \cite{mamakoukas2022robust,narayanan2023se }. These methods often demand large training datasets, leading to time-consuming and expensive data collection processes. Moreover, learned Koopman models often lack scalability beyond their training trajectories and fail to generalize across platforms, limiting their use for general-purpose modeling \cite{manaa2024koopman}.

To address the shortcomings of data-driven Koopman-based approaches, \cite{Chen2020SO3} proposed a data-free Koopman-linear realization on SO(3), which was later extended to quadrotor dynamics on SE(3) in \cite{zinage} with a linear time-invariant (LTI) quadrotor model with a very high dimensional virtual control input. However, \cite{zinage} neither investigates controller synthesis within this Koopman-linearized framework nor considers the curse of dimensionality that may arise in an LMPC scheme with such a high dimensional virtual control input. Moreover, applying the control requires solving an online least-squares problem, which challenges real-time computational reliability. Translating constraints from the virtual input to the actual input is also nontrivial and remains unaddressed.

Motivated by this gap, this letter introduces an alternative analytical (data-free) Koopman linear representation of quadrotor dynamics. The proposed formulation preserves the original dimension of the control vector in the lifted space, offering a more compact and tractable representation than \cite{zinage}. Furthermore, we demonstrate a controller synthesis procedure using this Koopman formulation, an aspect not explored in \cite{zinage}, and present the resulting LMPC framework, termed \emph{KQ-LMPC}, which aims to narrow the performance-complexity trade-off in quadrotor MPC design. The key contributions of this letter are:
(1) an analytical Koopman-linear representation of quadrotor dynamics that enables a compact lifted-space evolution, preserving the control input dimension of the full nonlinear model; (2) a formal proof of controllability for the derived model, validating its use for control;
(3) an LMPC scheme which leverages the Koopman-linear model to formulate a convex Quadratic Program (QP) that explicitly enforces both input and state constraints with robustness guarantee; and
(4) numerical and experimental validations of the proposed KQ-LMPC scheme demonstrating its real-world feasibility, superior computational efficiency, and tracking performance reasonably comparable to NMPC. To the best of our knowledge, this is the first data-free Koopman LMPC scheme to achieve experimental validation.
\vspace{-0.6em}

\section{Terminology}

Let $\mathbb{Z}^+$ denote the set of all positive integers. An inertial frame is defined as $\mc{I}_{\text{ref}}\triangleq\{ \overrightarrow{e_1},\overrightarrow{e_2},\overrightarrow{e_3} \}$, where $\overrightarrow{e_1}=[1,0,0]^\top$, $\overrightarrow{e_2}=[0,1,0]^\top$, and $\overrightarrow{e_3}=[0,0,1]^\top$, and a body-fixed frame is defined as $\mathfrak{B}_{\text{ref}}\triangleq\{ \overrightarrow{b_1},\overrightarrow{b_2},\overrightarrow{b_3} \}$, as depicted in \Cref{fig:quad_model}(a). The acceleration due to gravity is denoted by $g$ and we define $\bar{\mf{g}}=g\overrightarrow{e_3}=[0,0,g]^\top$. Let $\mf{I}_m$ denote the $m\times m$ identity matrix, $\mf{0}_m$ denote the $m \times m$ zero matrix,  $\mathbbm{1}_n$ denote the $n$-dimensional column vector of ones, and $\mf{0}_{m\times n}$  denote the $m \times n$ zero matrix. We let $\otimes$ denote the Kronecker product. The vectorization operation \( \operatorname{vec}(\cdot) : \mathbb{R}^{m \times n} \to \mathbb{R}^{mn} \) stacks the columns of a matrix \( \mathbf{B} \) into a vector $\underline{\mathbf{B}}$, while \( \operatorname{vec}^{-1}(\cdot) : \mathbb{R}^{mn} \to \mathbb{R}^{m \times n} \) recovers \( \mathbf{B} \) from $\underline{\mathbf{B}}$. Let $|a|$ denote the 2-norm of a vector $a$, and let $\| A\|$ denote the induced 2-norm of a matrix $A$.
For $\mathbf{p}\in\mathbb{R}^3$, the skew-symmetric matrix $\mathbf{p}^\times$ satisfies $\mathbf{p}\times\mathbf{q}=\mathbf{p}^\times\mathbf{q}$ for all $\mathbf{q}\in\mathbb{R}^3$, and the vee operator $(\cdot)^\vee$  recovers $\mathbf{p}$ from $\mathbf{p}^\times$, i.e., $(\mathbf{p}^\times)^\vee=\mathbf{p}$.
We define $\operatorname{blkdiag}\!\big(A_1, A_2, \dots, A_n \big)$ as the block diagonal matrix consisting of the matrices $A_1,\dots,A_n$. For a vector $\mathbf{e} \in \mathbb{R}^n$, we define the quadratic form as $\|\mathbf{e}\|_Q^2 = \mathbf{e}^\top Q \mathbf{e}$, with $Q \in \mathbb{R}^{n \times n}$ a positive (semi-) definite matrix. The special orthogonal group $SO(3)$ is defined as $\operatorname{SO}(3) := \{ A \in \mathbb{R}^{3\times 3} \mid A^\top A = \mathbf{I}_3,\; A A^\top = \mathbf{I}_3,\; \det(A)=1 \}$. Let $\mathbf{s} \in \mathbb{R}^3$ and $\mathbf{v} \in \mathbb{R}^3$ denote the position and linear velocity of the quadrotor's center of mass in $\mc{I}_{\text{ref}}$, respectively, and define $\mathbf{R} \in \operatorname{SO}(3)$ as the rotation matrix from $\mathfrak{B}_{\text{ref}}$ to $\mathcal{I}_{\text{ref}}$. Additionally, let $m$ denote the mass, 
$\mathbf{J} \in \mathbb{R}^{3\times3}$ the inertia matrix in $\mathfrak{B}_{\text{ref}}$, $\boldsymbol{\omega} \in \mathbb{R}^3$ the angular velocity in $\mathfrak{B}_{\text{ref}}$, $\boldsymbol{\tau} \in \mathbb{R}^3$ the total moment in $\mathfrak{B}_{\text{ref}}$, $f$ the total thrust along $\overrightarrow{b_3}$, $\mathbf{u} \in \mathbb{R}^4$ the control input vector defined as $\mathbf{u} = [f,\, \boldsymbol{\tau}^\top]^\top$, the state vector $\mathbf{x}\in\mathbb{R}^{18}$ defined as $\mathbf{x} = [\mathbf{s}^\top,\mathbf{v}^\top,\underline{\mathbf{R}}^\top,\boldsymbol{\omega}^\top]^\top$, and $\boldsymbol{\Omega}$ a skew symmetric matrix defined as $\boldsymbol{\Omega} = \boldsymbol{\omega}^\times$.


\vspace{-0.6em}

\section{Preliminaries} \label{sec:prelim}

\subsection{Quadrotor Dynamics}

\begin{figure}[!htpb]
\centering
\includegraphics[scale=0.28]{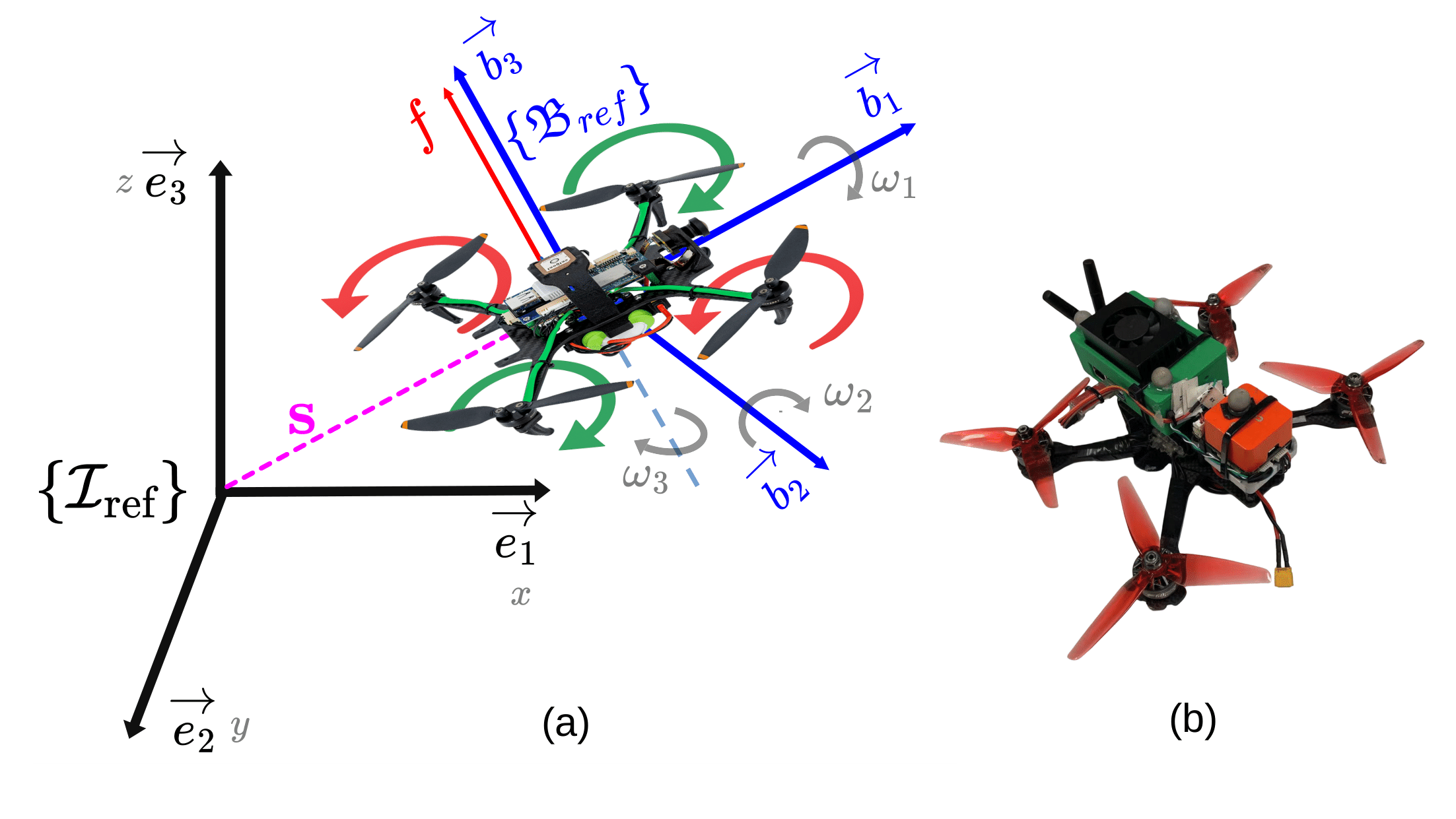}
\caption{(a) Quadrotor's coordinate frames, (b) The quadrotor used in the experiment. } 
\label{fig:quad_model}
\end{figure}

The governing equations of motion of a quadrotor, with the reference frames as depicted in Figure \ref{fig:quad_model}(a), can be described as follows \cite{lee_geom}:
\begin{equation}
    \Dot{\mathbf{s}} = \mathbf{v}, \
    \Dot{\mathbf{v}} = -\widetilde{\mathbf{g}}+\tfrac{f}{m}\mathbf{R}\overrightarrow{e_3}, \
    \dot{\mathbf{R}} = \mathbf{R}\boldsymbol{\Omega}, \
    \mathbf{J}\Dot{\boldsymbol{\omega}}= -\boldsymbol{\Omega} \mathbf{J} \boldsymbol{\omega} + \boldsymbol{\tau}.
\label{eq:quad_dynamics1}
\end{equation}
As shown in (\ref{eq:quad_dynamics1}), the quadrotor is an underactuated system, with \( f \) and \( \boldsymbol{\tau} \) as the control inputs. The set of all possible configurations of a quadrotor is described by the special Euclidean group \( \SE(3) \), which comprises the special orthogonal group \( \SO(3) \) and the position vector \( \mf{s} \). The dynamics in \eqref{eq:quad_dynamics1} can be represented in a more compact form as
\begin{equation}
    \Dot{\mathbf{s}} = \mathbf{v}, \
    \Dot{\mathbf{v}} = -\widetilde{\mathbf{g}}+(f/m)\mathbf{R}\overrightarrow{e_3}, \
    \dot{\mathbf{R}}  = \mathbf{R} \widehat{\boldsymbol{\Omega}}, \
    \dot{\boldsymbol{\omega}}= \hat{\mathbf{J}}\tilde{\boldsymbol{\tau}},
    \label{eq:quad_compact}
\end{equation}
where $\hat{\mathbf{J}} =  \mathbf{J}^{-1}$ and $\tilde{\boldsymbol{\tau}} = - \boldsymbol{\Omega}\mathbf{J} \widehat{\boldsymbol{\omega}} + \boldsymbol{\tau}$. To facilitate a control-affine representation, we introduce a modified control input $\widetilde{\mathbf{u}} = \mathscr{U}(\mathbf{x},\mathbf{u}) = [f,\, \tilde{\boldsymbol{\tau}}^\top]^\top$, such that the original input can be recovered via $\mathscr{U}^{-1}(\mathbf{x}, \widetilde{\mathbf{u}}) = \mathbf{u}$. With these definitions, we can write the control-affine form of \eqref{eq:quad_compact} as
\begin{equation}
    \dot{\mathbf{x}} = \mathbf{F}(\mathbf{x}, \mathbf{u}) = \mathbf{f(x)} + {\scriptstyle\sum_{i=1}^{4}}\mathbf{g}_i(\mathbf{x})\tilde{u}_i,
    \label{eq:quad_affine}
\end{equation}
with $\mathbf{f(x)} =   \big[ \mathbf{v}^\top, \ -\bar{\mf{g}}^\top, \ (\mathbf{R}\boldsymbol{\Omega})^\top, \ \mathbf{0}_{1\times3} \big]^\top$, 
 $\mathbf{g(x)} = \begin{bmatrix}
            \mathbf{0}_{1\times3} & (1/m) (\mathbf{R}\overrightarrow{e_3})^\top & \mathbf{0}_{1\times 9} & \mathbf{0}_{1\times 3} \\
            \mathbf{0}_{3\times3} & \mathbf{0}_{3\times3} & \mathbf{0}_{3\times9} & \hat{\mathbf{J}}
        \end{bmatrix}^\top,$
and $\mathbf{g}_i(\mathbf{x})$ is the $i$-th column of $\mathbf{g}(\mathbf{x})$ which corresponds to the $i$-th element $\tilde{u}_i$ of $\tilde{\mathbf{u}}$. Further details on the dynamics of the quadrotor model can be found in \cite{lee_geom}.

\subsection{Koopman Operator Theory}

The Koopman operator provides a functional framework for analyzing dynamical systems through observable functions on a Banach space \cite{koopman1931hamiltonian}. Consider a control-affine system 
\begin{equation}
    \dot{\mf{x}}(t) = \mathbf{F}(\mf{x},\mf{u}) = \mf{f}(\mf{x}) + {\scriptstyle\sum_{i=1}^{m}} \mf{g}_i(\mf{x})u_i,
    \label{eq:koopm_dyn}
\end{equation}
with state $\mf{x} \in \mc{M}\subset \mb{R}^n$, control $\mf{u} \in \mc{U}\subset \mb{R}^m$, drift $\mf{f}:\mc{M}\to \mc{M}$, and input vector fields $\mf{g}_i:\mc{M}\to \mc{M}$. Let $\mf\Phi_\mf{u}(t,\mf{x}_0)$ denote the flow of \eqref{eq:koopm_dyn} with initial condition $\mf{x}_0$. A measurable function $\varphi:\mc{M}\to\mb{R}$ is an observable, and $\mc{F}$ denotes a Banach space of continuously differentiable real-valued observables. The continuous-time Koopman (semi-)group $\mc{K}_\mf{u}^t:\mc{F}\to \mc{F}$ acts as 
\begin{equation}
 \mc{K}_\mf{u}^t\varphi(\mf{x}_0) = \varphi \circ \mf\Phi_\mf{u}(t,\mf{x}_0),
\label{eq:koop_comp}
\end{equation}
where $\circ$ denotes composition. Let $\hat{\varphi}(t)=\mc{K}_u^t\varphi(\mf{x}_0)$. Then  $\hat{\varphi}(t)$ is the solution to
$ \partial {\hat{\varphi}}/ \partial t = \mathbf{F} \cdot \nabla \hat{\varphi} := L_\mathbf{F} \hat{\varphi}$, with $\hat{\varphi}(0)=\varphi(\mf{x}_0)$. Here $\nabla$ is the gradient, $\cdot$ the inner product in $\mb R^n$, and $L_\mathbf{F}$ the infinitesimal generator of $\mc{K}_\mf{u}^t$ defined as $L_\mathbf{F}\varphi=\lim_{t\to0}\big[(\mc{K}_\mf{u}^t\varphi-\varphi)/t\big]$. Thus, the observable evolves as 
\begin{equation}
\label{eq:observable}
\dot{\varphi} = L_\mf{f}\varphi + {\scriptstyle\sum_{i=1}^m} L_{\mf{g}_i}\varphi\, u_i,
\end{equation}
where $L_\mf{f},L_{\mf{g}_i}$ are the Koopman generators for $\mf f$ and $\mf g_i$. Consider $\mathcal{F}$ to be the span of a countable collection of real-valued observables $\{\varphi_1(\cdot),\varphi_2(\cdot),\ldots\}$ with $\boldsymbol{\varphi}(\mf x)=[\varphi_1(\mf x),\varphi_2(\mf x),\ldots]^\top$. We define the lifted state $\mf z=\boldsymbol{\varphi}(\mf x)\in \mathcal{Z} \subset \mb R^{\aleph}$, where $\aleph$ is the cardinality. Let there exist a nonlinear reconstruction map $\boldsymbol{\varphi}^{-1}: \mathcal{Z}\to \mc{M}$ satisfying $\boldsymbol{\varphi}^{-1}(\mf z)=x$. The map $\boldsymbol{\varphi}^{-1}$ is a minimal representation which uses only the essential components of $z$ to reconstruct $x$. Assume that $\mc{F}$ is invariant under $L_{\mf{f}}$, i.e., $L_{\mf f}\mc{F}\subset \mc{F}$.
Under this assumption, the lifted dynamics admit a quasi-linear representation \cite{Surana2016}: 
\begin{equation}
    \dot{\mf z} = (\partial \boldsymbol{\varphi}(x)/\partial x)\mathbf{F}(\mf x, \mf u) = \mf{A z} + \mf{B}(\mf{z})\mf{u},
\end{equation}
where $\mf{Az} \! \! = \! L_\mf{f}\boldsymbol{\varphi}(\mf x)$ and $\mf B(\mf z)\mf u$ $ ={\scriptstyle\sum_{i=1}^m}L_{\mf{g}_i}\boldsymbol{\varphi}(x)\big|_{x=\boldsymbol{\varphi}^{-1}(\mf z)}u_i$. In practice, $\mf z$ is truncated to $\mf z=[\varphi_1(\mf x),\ldots,\varphi_\mc{N}(\mf x)]$ with $\mc{N} \gg n$, defined as $\mc{N} := \{ \mc N = \aleph \mid \aleph < \infty \}$, ideally ensuring vanishing residuals. However, constructing an invariant set of observables is nontrivial, as no universal procedure exists. 

\vspace{-0.6em}

\section{ Analytical Koopman Embedding for Quadrotor Dynamics
} \label{sec:dict}
We present a systematic approach for constructing the Koopman invariant lifted space \( \mathcal{F} \) for quadrotor dynamics \eqref{eq:quad_dynamics1} based on a proposed set of observables. The quadrotor dynamics \eqref{eq:quad_dynamics1} or \eqref{eq:quad_compact} can be categorized into three distinct interdependent components: (i) attitude dynamics, governing the evolution of \( \mathbf{R} \) and \( \boldsymbol{\omega} \); (ii) position dynamics, describing the evolution of \( \mf{s} \); and (iii) linear velocity dynamics, which includes the effects of gravitational acceleration and total thrust on \( \mf{v} \). To systematically capture these dynamics in the lifted space \( \mathcal{F} \), we propose a collection of observables that encode these components. Specifically, we define four sets of observations: \( \mathbf{p}_k\), associated with position; \( \mathbf{y}_k \), associated with linear velocity; \( \mathbf{h}_k \), associated with gravitational acceleration; and \( \mathbf{z}_j \), associated with attitude dynamics. For $k,\ j \in \mathbb{Z}^+$,  we propose the following observables 
\begin{equation}
\begin{aligned}\label{eq:obs}
\textbf{p}_k(\mathbf{x}) &= (\boldsymbol{\Omega}^\top)^{k-1}\mathbf{R}^\top \mf{s}, &
\textbf{y}_k(\mathbf{x}) &= (\boldsymbol{\Omega}^\top)^{k-1}\mathbf{R}^\top\mathbf{v}, \\
\textbf{h}_k(\mathbf{x}) &= -(\boldsymbol{\Omega}^\top)^{k-1}\mathbf{R}^\top \bar{\mf{g}}, &
\underline{\mathbf{z}}_j(\mathbf{x}) &= \operatorname{vec}(\mathbf{R} \boldsymbol{\Omega}^{j-1}),
\end{aligned}
\end{equation}
with the recurrence relations for $k,\ j >1$ given by 
\begin{equation}
\begin{aligned}  
\mf{p}_{k+1}(\mathbf{x}) &= \boldsymbol{\Omega}^\top \mf{p}_{k}(\mathbf{x}), \ \ 
\mf{y}_{k+1}(\mathbf{x}) = \boldsymbol{\Omega}^\top \mf{y}_{k}(\mathbf{x}) , \\
\mf{h}_{k+1} (\mathbf{x}) &= \boldsymbol{\Omega}^\top \mf{h}_{k}(\mathbf{x}) , \ \ 
\underline{\mathbf{z}}_{j+1}(\mathbf{x}) = \operatorname{vec}\big(\mathbf{z}_{j}(\mathbf{x}) \boldsymbol{\Omega} \big).
\end{aligned}
\label{eq:rec_rel}
\end{equation}

\begin{remark}
From \eqref{eq:obs}, it is worth noting that $\mathbf{z}_1 = \mathbf{R}$ or $\underline{\mathbf{z}}_1 = \operatorname{vec}(\mathbf{R})$, $\mathbf{s} = \mathbf{z}_1\mathbf{p}_1$, $\mathbf{v} = \mathbf{z}_1\mathbf{y}_1$, $\mathbf{h}_1 = -\mathbf{z}_1^\top \widetilde{\mathbf{g}}$, $\boldsymbol{\Omega} = \mathbf{z}_1^\top \mathbf{z}_2$ and  $\boldsymbol{\omega} = (\mathbf{z}_1^\top \mathbf{z}_2)^\vee$.
\end{remark}

\begin{assumption}
\label{assum:control_limits}
The control input $\mathbf{u}$ is constrained to the set $\mathcal{U} := \{ \mathbf{u} \in \mathbb{R}^4 \mid \mathbf{u}_{\min} \leq \mathbf{u} \leq \mathbf{u}_{\max} \}$, with inequalities interpreted component-wise. Consequently, the system state evolves within a compact set $\boldsymbol{\mathcal{X}} := \{ \mathbf{x} \in \mathbb{R}^{18} \mid \boldsymbol{\omega}_{lb} \leq \boldsymbol{\omega} \leq \boldsymbol{\omega}_{ub},  \ \mathbf{s}_{lb} \leq \mathbf{s} \leq \mathbf{s}_{ub},\ \mathbf{v}_{lb}\leq \mathbf{v} \leq \mathbf{v}_{ub},\ \mathbf{R} \in \mathrm{SO}(3)\}$, where the inequalities are considered point-wise. 
Let these inequalities imply that
$
|\boldsymbol{\omega}| \leq \omega_{\max}$,  $ |\mathbf{s}| \leq s_{\max}$, and $|\mathbf{v}| \leq v_{\max},
$
for finite constants $s_{\max}, v_{\max}, \omega_{\max} \in \mathbb{R}_{>0}$, with $\omega_{\max} < 1/\sqrt{2}$. These limits ensure that the quadrotor operates within feasible physical and stability limits.
\end{assumption}

\subsection{Invariant Observables and Lifted Quasilinear Realization}

We define a countably infinite collection of observables proposed in \eqref{eq:obs} as 
\begin{equation}
   \!\!\!  \varphi(\mathbf{x})   = \big\{ \{p_{k_i}\}_{\mathfrak{i}=1}^3,\{y_{k_i}\}_{\mathfrak{i}=1}^3, \{h_{k_i}\}_{\mathfrak{i}=1}^3, \{z_{j_i}\}_{\mathfrak{i}=1}^9  \big\} _{k,j=1}^{\infty},
    \label{eq;superc}
\end{equation}
where  $\mathbf{p}_k\!=\![p_{k_1}, p_{k_2}, p_{k_3}]^\top$, $\mathbf{y}_k\!=\![y_{k_1}, y_{k_2}, y_{k_3}]^\top\!$, $\mathbf{h}_k\!=\![h_{k_1}, h_{k_2}, h_{k_3}]^\top\!$, and $\mathbf{\underline{z}}_j\!=\![z_{j_1}, \dots, z_{j_9}]^\top\!$ (representing individual scalar components extracted from the vector form in \eqref{eq:obs}). These observables belong to the function space $\mathcal{F}$, referred to as the \emph{lifted space} associated with quadrotor dynamics.
\begin{lemma}
Under Assumption~\ref{assum:control_limits}, $\mathcal{F}\triangleq \overline{\mathrm{span}\{\varphi(\cdot)\}}$, i.e., the closure of the finite linear combinations of the components of $\varphi$ is a Hilbert space with respect to $L^2$ norm, and hence it is also a Banach space.
\end{lemma}
\begin{proof}
     Each observable in the collection $\varphi$ are $C^1$ and bounded on $\boldsymbol{\mathcal{X}}$ due to compactness. Let $L^2\triangleq L^2_\mu(\boldsymbol{\mathcal{X}})$ be the space of square Lebesgue-integrable functions on the compact (and therefore finite-measure) set $\boldsymbol{\mathcal{X}} \in\mb R^{18}$. $L^2$ is a Hilbert space with an associated inner product
     \[\langle f,g \rangle \triangleq \int\limits_{\boldsymbol{\mathcal{X}}}f(\cdot)g(\cdot)d\mu\ .\]
By construction, each component of $\varphi(\cdot)$ is bounded on $\boldsymbol{\mathcal{X}}$, i.e., $\spanop \{\varphi(\cdot)\}\subset L^\infty_\mu(\boldsymbol{\mathcal{X}})$. This fact, with $\boldsymbol{\mathcal{X}}$ being a finite measure set, yields $\spanop\{\varphi(\cdot)\}\subset L^2$. Let $\mc F \triangleq \overline{\spanop\{\varphi\}}$ where the closure is taken in $L^2$ with the standard 2-norm induced by its inner product. The set $\mc{F}$ is essentially a closure of all finite linear combinations of the observables in $\varphi(\cdot)$. Therefore, $\mc{F}$ is a linear subspace of $L^2$ by construction and it is closed by the definition of \emph{closure}. The space $L^2$ is a Hilbert space with the standard induced 2-norm $\norm{f}_2 = \sqrt{\langle f, f \rangle}$. To show $\mc{F}$ is Hilbert, it suffices to show that $\mc{F}$ is complete in the induced norm. Let $\{f_k\}$ be a Cauchy sequence in $\mc{F}$. Then $\{\phi_k\}$ is Cauchy in $L^{2}$ as well, hence converges in $L^{2}$ to some $\phi^*\in L^{2}$. Because $\mc{F}$ is closed in $L^{2}$, the limit $\phi^*$ must belong to $\mc{F}$. Therefore, every Cauchy sequence in $\mc{F}$ converges to an element of $\mc{F}$, so $\mc{F}$ is complete. Thus, $\mc{F}$, equipped with the restriction of $\langle\cdot,\cdot\rangle$ from $L^{2}$, is a Hilbert space. This completes the proof.
\end{proof}

\begin{theorem}
\label{Thm:1}
The lifted space $\mathcal{F} := \overline{\mathrm{span}\{\varphi(\cdot)\}}$, with $\varphi$ as defined in \eqref{eq;superc}, is Koopman-invariant, i.e., invariant under $L_{\mf f}$ for the dynamics \eqref{eq:quad_affine} of the quadrotor.
\end{theorem}
\begin{proof}
Based on \eqref{eq:quad_compact}, \eqref{eq:obs}, and \eqref{eq:rec_rel}, consider the individual evolution of the constituent observables of $\varphi$ for $k, j = 1$, expressed as follows: 
\begin{equation}
\begin{aligned}
 \!\!\! \!\!  \Dot{\mathbf{p}}_1 \!= \!\mathbf{p}_2+\mathbf{y}_1,\ 
    \Dot{\mathbf{y}}_1 \! =\! \mathbf{y}_2 \!+ \!\mathbf{h}_1 \!+\!f\overrightarrow{e_3}/m,\ 
    \Dot{\mathbf{h}}_1 \!=\!  \mathbf{h}_2,\  \Dot{\underline{\mathbf{z}}}_1 \! = \! \underline{\mathbf{z}}_2.
    \label{eq:obs_der1}
\end{aligned}
\end{equation}
Furthermore, for $k, j > 1$, we have
\begin{equation}
\begin{aligned}
 \dot{\mathbf{p}}_k &= \! \mathbf{p}_{k+1} \!+\! \mathbf{y}_k \!+ \! {\scriptstyle\sum_{i=1}^{k-1}} \mathfrak{P} \mathbf{p}_1, \ \dot{\mathbf{h}}_k = \mathbf{h}_{k+1} \!+ \! {\scriptstyle\sum_{i=1}^{k-1}} \mathfrak{P} \mathbf{h}_1,\\ 
\Dot{\mathbf{y}}_k &= \mathbf{y}_{k+1} + \mathbf{h}_k + (\boldsymbol{\Omega}^\top)^{\,k-1}\tfrac{f}{m}\overrightarrow{e_3} + {\scriptstyle\sum_{i=1}^{k-1}} \mathfrak{P} \mathbf{y}_1, \\
\Dot{\underline{\mathbf{z}}}_j &= \underline{\mathbf{z}}_{j+1} + \operatorname{vec}\bigr( \mathbf{R}\ {\scriptstyle\sum_{i=1}^{j-1}}(\boldsymbol{\Omega})^{\,i-1} \big(\hat{\mathbf{J}} \tilde{\boldsymbol{\tau}}\big)^\times (\boldsymbol{\Omega})^{\,j-1-i} \bigr),
\label{eq:obs_der2}
\end{aligned}
\end{equation}
where $\mathfrak{P} =  (\boldsymbol{\Omega}^\top)^{\,i-1} \big(\hat{\mathbf{J}} \tilde{\boldsymbol{\tau}}\big)^{\times^\top} (\boldsymbol{\Omega}^\top)^{\,k-1-i}$. 
Let $\boldsymbol{\varphi}: \boldsymbol{\mathcal{X}} \to \boldsymbol{\mathcal{Z}} \subset \mathbb{R}^\aleph$ be a nonlinear mapping defined as
$\boldsymbol{\varphi}(\mathbf{x}) = \mathbf{X}$, with
$\boldsymbol{\mathcal{Z}} :=  \boldsymbol{\varphi}(\mathcal{X}) = \{  \mathbf{X} \in \mathbb{R}^\aleph   \mid \  \mathbf{x} \in \boldsymbol{\mathcal{X}} \}$, where the lifted state $\mathbf{X}$ is
\begin{align}
        \mathbf{X} = [
          \mathbf{p}^\top_1 \dots  \mathbf{p}^\top_M \ \mathbf{y}^\top_1 \dots \mathbf{y}^\top_M  \ \mathbf{h}^\top_1 \dots \mathbf{h}^\top_M \ \underline{\mathbf{z}}^\top_1 \ \dots \ \underline{\mathbf{z}}^\top_{N}]^\top,
        \label{eq:define_X}
\end{align}
with the cardinality $\aleph = 9M+9N$. Here, $M,N \to \infty$ for $k = \{1,2,\dots,M\}$ and $j\in \{1,2,\dots,N\}$.
Further, from \eqref{eq:obs}, the reconstruction map
$\boldsymbol{\varphi}^{-1}:\boldsymbol{\mathcal{Z}} \to \boldsymbol{\mathcal{X}}$ , defined as $\boldsymbol{\varphi}^{-1}(\mathbf{X})=\mathbf{x}$, can be explicitly expressed as
\begin{equation}
    \! \! \boldsymbol{\varphi}^{-1}(\mathbf{X}) = 
        \bigl[{\scriptstyle(\mathbf{z}_1 \mathbf{p}_1\bigr)^\top,  \bigl(\mathbf{z}_1 \mathbf{y}_1\bigr)^\top, \underline{\mathbf{z}}_1^\top, \bigl(\mathbf{z}_1^\top\,\mathbf{z}_2\bigr)^{\vee^\top}}\bigr]^\top
        \label{eq:Dx}
\end{equation}
The map $\boldsymbol{\varphi}^{-1}$ depends only on the minimal set of components in $\mathbf{X}$ required to uniquely reconstruct $\mathbf{x}$. Under the change of co-ordinates $\boldsymbol{\varphi}$, it follows from \eqref{eq:quad_compact}, \eqref{eq:quad_affine}, \eqref{eq:obs_der1}, and \eqref{eq:obs_der2}, that
\begin{align}
        \Dot{\mathbf{X}} = \mathbf{AX} + \boldsymbol{\mathcal{B}}(\mathbf{X})\widetilde{\mathbf{u}}, 
        \label{eq:lin_calB_eq}
\end{align} 
where $\mathbf{A} \in \mathbb{R}^{\aleph \times \aleph}$ and $\boldsymbol{\mathcal{B}} \in \mathbb{R}^{\aleph \times 4}$, defined as 
\begin{equation}
\begin{aligned}
\mathbf{A} =&   \operatorname{blkdiag} \big( \mathbf{A}_p,  \mathbf{A}_a \big), \
     \boldsymbol{\mathcal{B}} \! = \! \big[
        \mathbf{B}^\top_1 \ \mathbf{B}^\top_2 \  \dots \ \mathbf{B}^\top_{3M+N}
    \big]^\top\!\!\!, \\
    \mathbf{A}_p &= \bigl 
        [\mathbf{A}_1 , \mathbf{0}_{3M} ;  \
        \mathbf{0}_{3M} , \mathbf{A}_2; \
        \mathbf{0}_{3M \times 6M} , \mathbf{A}_3
    \bigr ], \\ 
    \mathbf{A}_1 &= \mathbf{A}_2 = [
        \mathcal{A}_1, \ \mathcal{A}_2], \\
    \mathbf{A}_a &=  [
        \mathbf{0}_{9(N-1) \times 9} , \mathbf{I}_{9(N-1)} ; \
        \mathbf{0}_{9 \times 9} , \mathbf{0}_{9 \times 9(N-1)}],  \\
    \mathcal{A}_1 &=  [
        \mathbf{0}_{3(M-1) \times 3},\  \mathbf{I}_{3(M-1)} ; \
        \mathbf{0}_3, \ \mathbf{0}_{3 \times 3(M-1)}
    ], \\
    \mathcal{A}_2 &=  [
        \mathbf{I}_{3(M-1)},\ \mathbf{0}_{3(M-1) \times 3}; \
        \mathbf{0}_{3 \times 3(M-1)},\ \mathbf{0}_3
    ],\\
    \mathbf{A}_3 &=  [
        \mathbf{0}_{3(M-1) \times 3} ,\ \mathbf{I}_{3(M-1)} ;\
        \mathbf{0}_3 ,\ \mathbf{0}_{3 \times 3(M-1)}
    ],\\
 \mathbf{B}_{\mathfrak{k}} &= \begin{cases} 
    {\scriptstyle\mathbf{0}_{3 \times 4}}, \!\!\!& \scriptstyle \text{if } \mathfrak{k} = 1, \\
    {\scriptstyle[\mathbf{0}_{3 \times 1},\, \boldsymbol{\Psi}^{\mathfrak{k}}(\mathbf{p}_1)]}, & \scriptstyle \text{if } 2 \leq \mathfrak{k} \leq M, \\
    {\scriptstyle[\overrightarrow{e_3}/m,\, \mathbf{0}_3]}, \!\!\!& \scriptstyle \text{if } \mathfrak{k} = M+1, \\
    {\scriptstyle[\boldsymbol{\zeta}^{\mathfrak{k}-M-1},\, \boldsymbol{\Psi}^{\mathfrak{k}-M}(\mathbf{y}_1)]}, \!\!\!& \scriptstyle \text{if } M+2 \leq \mathfrak{k} \leq 2M, \\
    {\scriptstyle\mathbf{0}_{3\times 4}}, \!\!\!& \scriptstyle \text{if } \mathfrak{k} = 2M+1, \\
    {\scriptstyle[\mathbf{0}_{3 \times 1},\, \boldsymbol{\Psi}^{\mathfrak{k}-2M}(\mathbf{h}_1)]}, & \scriptstyle \text{if } 2M+2 \leq \mathfrak{k} \leq 3M, \\
    {\scriptstyle\mathbf{0}_{9\times 4}}, \!\!\!& \scriptstyle \text{if } \mathfrak{k} = 3M+1, \\
    {\scriptstyle[\mathbf{0}_{9 \times 1},\, (\mathbf{I}_3 \otimes \mathbf{z}_1)\mathbf{G}^{\mathfrak{k}-3M}]}, \!\!\!& \scriptstyle \text{if } \mathfrak{k} > 3M+1.
\end{cases},
\label{eq:define_calB}
\end{aligned}
\end{equation} 
with $\mathfrak{k} \in \{ 1, 2, \dots, 4M \}$, and 
\begin{equation}
\begin{aligned}
        \boldsymbol{\Psi}^{\mathfrak{k}}(\cdot) &=  {\scriptstyle\sum_{i=1}^{\mathfrak{k}-1}}(\mathbf{z}_2^\top\mathbf{z}_1)^{i-1} \bigl( (\mathbf{z}_2^\top\mathbf{z}_1)^{\mathfrak{k}-1-i} \cdot\bigr)^\times \hat{\mathbf{J}}, \\
        \mathbf{g}^{\mathfrak{k}}_\mathfrak{q} &= \operatorname{vec} \big( {\scriptstyle\sum_{i=1}^{\mathfrak{k}-1}} (\mathbf{z}_1^\top\mathbf{z}_2)^{i-1} \mathbf{j}_\mathfrak{q}^\times (\mathbf{z}_1^\top\mathbf{z}_2)^{\mathfrak{k}-1-i} \big),\\
        \mathbf{G}^{\mathfrak{k}} &=\begin{bmatrix}
        \mathbf{g}^{\mathfrak{k}}_1 & \mathbf{g}^{\mathfrak{k}}_2 & \mathbf{g}^{\mathfrak{k}}_3
    \end{bmatrix},\ \boldsymbol{\zeta}^{\mathfrak{k}} = (\boldsymbol{\Omega}^\top)^{\mathfrak{k}}\tfrac{\overrightarrow{e_3}}{m},
\end{aligned}
\end{equation}
where $\mathfrak{q}\in\{1,2,3\}$ and $\mathbf{j}_\mathfrak{q}$ represents $\mathfrak{q}$-th column of $\hat{\mathbf{J}}$.
Under the action of the Koopman generator, the dynamics of the quadrotor in the lifted-space becomes
\begin{equation}
\begin{aligned}
    \dot{\mathbf{X}} &= \tfrac{\partial \boldsymbol{\varphi}(\mf{x})}{\partial \mf{x}} \big( \mathbf{f(x)} + {\scriptstyle\sum_{i=1}^{4}}\mathbf{g}_i(\mathbf{x})\widetilde{u}_i \big), \\
    & = L_{\mf{f}} \boldsymbol{\varphi}(\mathbf{x}) +  {\scriptstyle\sum_{i=1}^{4}}L_{\mf{g}i} \boldsymbol{\varphi}(\mf{x}) \big|_{\mf{x} = \boldsymbol{\varphi}^{-1}( \mf{X})} \widetilde{u}_i
\end{aligned}
\end{equation}
It follows apparently from \eqref{eq:quad_affine}, \eqref{eq:lin_calB_eq}, and \eqref{eq:Dx} that
\begin{equation}
    \mathbf{AX} =  L_{\mf{f}} \boldsymbol{\varphi}(\mathbf{x}), \ \boldsymbol{\mathcal{B}}\mathbf{(X)}\widetilde{\mathbf{u}} =  
    {\scriptstyle\sum_{i=1}^{4}} L_{\mf{g}i} \boldsymbol{\varphi}(x) \big|_{\mf{x} = \boldsymbol{\varphi}^{-1}( \mf{X})} \widetilde{u}_i.
\end{equation}
Therefore, the countably infinite supercollection $\boldsymbol{\varphi}$, introduced in \eqref{eq;superc}, spans the Koopman-invariant lifted space $\mathcal{F}$ associated with the quadrotor dynamics, and \eqref{eq:lin_calB_eq} is the associated quasilinear form of quadrotor dynamics under the action of the Koopman generator.
\end{proof}

\subsection{Finite-Dimensional Representation and Controllability}

The Koopman operator provides a globally linear representation in an infinite-dimensional Banach space of observables, as defined for quadrotor dynamics in \eqref{eq:lin_calB_eq}. However, practical implementation necessitates a finite-dimensional approximation to ensure tractability. Consequently, \eqref{eq:lin_calB_eq} must be truncated to \(M,N < \infty\) to obtain a finite-dimensional representation. It is important to ensure that the discarded terms remain negligibly small. In particular, if the observables and their time derivatives vanish asymptotically (pointwise) as \(M,N \to \infty\), the resulting finite-dimensional approximation is valid~\cite{zinage}.

\begin{theorem}
    Under \Cref{assum:control_limits}, the sequences of observables ${ \mathbf{p}_k, \mathbf{y}_k, \mathbf{h}_k, \underline{\mathbf{z}}_j }$ exhibit a pointwise contraction in $k$ and $j$, and decay exponentially to zero as $k,j \rightarrow \infty$. Furthermore, the derivatives $\dot{\mathbf{p}}_k, \dot{\mathbf{y}}_k, \dot{\mathbf{h}}_k$, $\dot{\underline{\mathbf{z}}}_j$ also converge pointwise and exponentially to zero as $k,j \rightarrow \infty$.
    \label{thm2}
\end{theorem}
\begin{proof}
    By the submultiplicative property of both the 2-norm and the Frobenius norm, the recurrence relation \eqref{eq:rec_rel} yields
\begin{equation}
    \begin{aligned}
        |\mathbf{p}_{k+1}| &\leq \omega_{\max} |\mathbf{p}_k|,\quad 
        |\mathbf{y}_{k+1}| \leq \omega_{\max} |\mathbf{y}_k|, \\
        |\mathbf{h}_{k+1}| &\leq \omega_{\max} |\mathbf{h}_k|,\quad 
        |\underline{\mathbf{z}}_{j+1}| \leq \sqrt{2}\ \omega_{\max} |\underline{\mathbf{z}}_j|.
    \end{aligned}
\end{equation}
Since $\omega_{\max} < 1/\sqrt{2}$, the sequences of observables satisfy a pointwise contraction property with respect to $k$ and $j$. Moreover, \eqref{eq:obs} gives
    \begin{equation}
        \begin{aligned}
            |\mathbf{p}_k| &\leq \omega_{\max}^{k-1} s_{\max},\quad 
            |\mathbf{y}_k| \leq \omega_{\max}^{k-1} v_{\max}, \\
            |\mathbf{h}_k| &\leq \omega_{\max}^{k-1} g,\quad 
            |\underline{\mathbf{z}}_j| \leq \sqrt{3} \left(\sqrt{2}\ \omega_{\max}\right)^{j-1},
        \end{aligned}
\end{equation}
justifying exponential decay of the norm with increasing $k$ and $j$, as $\omega_{\max} < 1/\sqrt{2}$, i.e., $ \lim_{k\to\infty} |\mf{p}_k| = \lim_{k\to\infty} |\mf{y}_k| = \lim_{k\to\infty} |\mf{h}_k| = \lim_{j\to\infty} | \underline{\mf{z}}_j| = 0$. Since the sequences of observables are finite dimensional, norm convergence implies componentwise convergence. Therefore, 
    \begin{equation}
        \lim_{k\to\infty} \mf{p}_k = \lim_{k\to\infty} \mf{h}_k = \lim_{k\to\infty} \mf{p}_k = \lim_{j\to\infty} \underline{\mf{z}}_j = \mathbf{0}.
        \label{eq:obs_decay}
    \end{equation}
    Since the control action is bounded, assume $|\tilde{\boldsymbol{\tau}}| \leq c_{3} < \infty $, $|f| \leq c_4 < \infty$, $(c_4 \overrightarrow{e_3} / m)=c_5$, $\|\hat{\mathbf{J}}\| = c_6$, and $c_3c_6 = \gamma_1$. It follows from \eqref{eq:obs_der2} that 
\begin{equation}
    \begin{aligned}
      \! \! \! \!    |\Dot{\mathbf{p}}_{k}| \leq & |\mathbf{p}_{k+1}| + |\mathbf{y}_{k}| + (k-1)\,s_{\max}\,\gamma_1\,\omega_{\max}^{\,k-2},\\
       \! \! \! \!  |\Dot{\mathbf{y}}_{k}| \leq & |\mathbf{y}_{k+1}| + |\mathbf{h}_{k}| + c_5 \omega_{\max}^{\,k-1} + (k-1)\, v_{\max}\,\gamma_1\,\omega_{\max}^{\,k-2},\\
      \! \! \! \!   |\Dot{\mathbf{h}}_{k}| \leq &  |\mathbf{h}_{k+1}| + (k-1)\,g\,\gamma_1\,\omega_{\max}^{\,k-2},\\
     \! \! \! \!   |\Dot{\underline{\mathbf{z}}}_{j}| \leq & |\underline{\mathbf{z}}_{j+1}| + (j-1)\sqrt{3}\,\sqrt{2}^{\,j-1}\,\gamma_1\,\omega_{\max}^{\,j-2}.
        \label{eq:obs_der_decay}
    \end{aligned}
\end{equation}
Therefore, Using the results from \eqref{eq:obs_decay}, the norms of the derivatives \eqref{eq:obs_der_decay} decay exponentially to zero, as $\omega_{\max} < 1/\sqrt{2}$, i.e.,  $\lim_{k\to\infty} |\dot{\mf{p}}_k| = \lim_{k\to\infty} |\dot{\mf{y}}_k| = \lim_{k\to\infty} |\dot{\mf{h}}_k| = \lim_{j\to\infty} |\dot{\underline{\mf{z}}}_j| = 0$. Here, the norm convergence implies pointwise convergence and gives 
\begin{equation}
\lim_{k\to\infty} \dot{\mf{p}}_k = \lim_{k\to\infty} \dot{\mf{y}}_k = \lim_{k\to\infty} \dot{\mf{h}}_k = \lim_{j\to\infty} \dot{\underline{\mf{z}}}_j = \mathbf{0}.
\end{equation}
    This completes the proof.
\end{proof}

\begin{remark} \textbf{Relaxing $\omega_{\max}<1/\sqrt{2}$ condition:} Consider a scaling factor $\mathfrak{w} \!>\! \sqrt{2}\cdot\omega_{\max}$ for $\omega_{\max} \! =\! \underset{\omega}{\max}(|\vect{\omega}|)$,
and the scaled angular velocity defined as
$\widehat{\boldsymbol{\omega}} = \boldsymbol{\omega}/\mathfrak{w}$, with the corresponding skew-symmetric matrix  
$\widehat{\boldsymbol{\Omega}} = \widehat{\boldsymbol{\omega}}^\times$. Under this scaling of $\boldsymbol{\omega}$, \eqref{eq:quad_compact} becomes
\begin{equation}
    \Dot{\mathbf{s}} = \mathbf{v}, \
    \Dot{\mathbf{v}} = -\widetilde{\mathbf{g}}+(f/m)\mathbf{R}\overrightarrow{e_3}, \
    \dot{\mathbf{R}}  = \mathfrak{w} \mathbf{R} \widehat{\boldsymbol{\Omega}}, \
    \dot{\widehat{\boldsymbol{\omega}}}= \hat{\mathbf{J}}\hat{\boldsymbol{\tau}},
    \label{eq:quad_compact2}
\end{equation}
where $\tilde{\boldsymbol{\tau}} = - \mathfrak{w}\widehat{\boldsymbol{\Omega}}\mathbf{J} \widehat{\boldsymbol{\omega}} + \boldsymbol{\tau} / \mathfrak{w}$. With the new state vector $\mathbf{x} = [\mathbf{s}^\top,\mathbf{v}^\top,\underline{\mathbf{R}}^\top,\widehat{\boldsymbol{\omega}}^\top]^\top$, we redefine \eqref{eq:obs} and \eqref{eq:rec_rel} by replacing $\boldsymbol{\Omega}$ with $\widehat{\boldsymbol{\Omega}}$. Therefore, we can add to \Cref{assum:control_limits} that $|\widehat{\boldsymbol{\omega}}|\leq\bar{\omega}$, where $\bar{\omega}<1/\sqrt{2}$ holds for any finite $\omega_{\max}$. With these modifications, it follows that \Cref{Thm:1} holds with system matrix $\widehat{\mathbf{A}}=\mathfrak{w}\mathbf{A}$. By the same reasoning, \Cref{thm2} is also valid. As the arguments mirror those in the original setting, the proofs are omitted for brevity.
\end{remark}

\begin{remark}
By Theorem \ref{thm2}, the infinite-dimensional lifted dynamics in \eqref{eq:lin_calB_eq} admit a finite-dimensional truncation with \( \mc{N} = \aleph \big|_{M,N < \infty} \),  \( \mathbf{X} \in \bar{\boldsymbol{\mathcal{Z}}} \subset \ \mathbb{R}^\mc{N} \), \( \mathbf{A} \in \mathbb{R}^{\mc{N} \times \mc{N}} \), \( \boldsymbol{\mathcal{B}} \in \mathbb{R}^{\mc{N} \times 4} \), and \( \mc{N} \gg 18 \). While maintaining notational consistency, the truncation yields a finite-dimensional quasilinear representation of the quadrotor dynamics given as 
    \begin{equation}
        \Dot{\mathbf{X}} = \mathbf{A}\mathbf{X} + \boldsymbol{\mathcal{B}}(\mathbf{X})\widetilde{\mathbf{u}}, 
        \label{eq:lin_calB_eq_finite}
    \end{equation}
    where \( \mathbf{X} \)  and (\( \mathbf{A}, \boldsymbol{\mathcal{B}} \)) inherit their definitions from \eqref{eq:define_X} and \eqref{eq:define_calB}, respectively. Furthermore, $\boldsymbol{\varphi}^{-1}:\bar{\boldsymbol{\mathcal{Z}}} \to \boldsymbol{\mathcal{X}}$ inherits its definition from \eqref{eq:Dx}, and $\bar{\boldsymbol{\mathcal{Z}}} := \boldsymbol{\varphi}(\boldsymbol{\mathcal{X}}) = \{  \mathbf{X} \in \mathbb{R}^\mathcal{N}   \mid \  \mathbf{x} \in \boldsymbol{\mathcal{X}} \}$.
\end{remark}

\begin{proposition}
    The quasilinear form of quadrotor dynamics in lifted space \eqref{eq:lin_calB_eq_finite}, which is an approximation of \eqref{eq:lin_calB_eq}, can be equivalently interpreted as a linear time-invariant (LTI) system with a lifted state-dependent control input.
\end{proposition}
\begin{proof}
    Let $\tilde{\boldsymbol{\mathcal{B}}} \in \mathbb{R}^{(\mc{N}-17) \times 4}$ denote the reduced matrix obtained by removing all zero rows from $\boldsymbol{\mathcal{B}}$. From \eqref{eq:define_calB}, it follows that
\begin{align}
  \begin{split}
       \!\!\!\! \tilde{\boldsymbol{\mathcal{B}}} = {\scriptstyle\big[ \mathbf{B}_2^\top, \dots ,\ \mathbf{B}_M^\top, \ \bar{\mathbf{B}}_{M+1}^\top, \dots,\ \mathbf{B}_{2M}^\top, \mathbf{B}_{2M+2}^\top,   \ , \dots, \ \mathbf{B}_{ 3M}^\top,} \\
       {\scriptstyle\ \ \quad \mathbf{B}_{ 3M+2}^\top, \dots\!,\ \mathbf{B}_{3M+N}^\top \big]^\top} \!\!\!,
  \end{split}
\end{align}
    where $\bar{\mathbf{B}}_{M+1} = [1/m, \ \mathbf{0}_{1\times 3}]$. Define the lifted state-dependent control $\mathbf{U} \in \mathbb{R}^{\mc{N}-17}$ as
    \begin{equation}
        \mathbf{U} = \tilde{\boldsymbol{\mathcal{B}}} \widetilde{\mathbf{u}},
    \text{ such that } \boldsymbol{\mathcal{B}} = \bar{\boldsymbol{\mathcal{B}}}\tilde{\boldsymbol{\mathcal{B}}}, \text{ and } 
        \boldsymbol{\mathcal{B}}\widetilde{\mathbf{u}} = \bar{\boldsymbol{\mathcal{B}}}\mathbf{U}, \label{eq:least_sq}
    \end{equation}
    where $\bar{\boldsymbol{\mathcal{B}}} \!\in\! \mathbb{R}^{\mc{N} \times (\mc{N}-17)\!}$ is a matrix of constant entries given as
\begin{equation}
\begin{aligned}
    \bar{\boldsymbol{\mathcal{B}}} =& \operatorname{blkdiag} \big(\mathcal{B}_1,\mathcal{B}_2,\mathcal{B}_3,\mathcal{B}_4  \big), \text{ with }\\
\mathcal{B}_1, \mathcal{B}_3 
&= \bigl[\mathbf{0}_{\scriptstyle 3(M-1)\times 3},\;\mathbf{I}_{\scriptstyle 3(M-1)} \bigr]^\top,\\
\mathcal{B}_2 
&= [
\overrightarrow{e_3} 
    ,\ \mathbf{0}_{\scriptstyle 3 \times 3(M-1)}; \ 
\mathbf{0}_{\scriptstyle 3(M-1)\times 1} 
    ,\ \mathbf{I}_{\scriptstyle 3(M-1)}
],\\
\mathcal{B}_4 
&= \bigl[\mathbf{0}_{\scriptstyle 9(N-1)\times 9},\;\mathbf{I}_{\scriptstyle 9(N-1)}\bigr]^\top.
\end{aligned}
\end{equation}
Substituting $\bar{\mathcal{B}}\mathbf{U}$ into \eqref{eq:lin_calB_eq_finite} yields the LTI system
\begin{equation}
    \Dot{\mathbf{X}} = \mathbf{A}\mathbf{X} + \bar{\boldsymbol{\mathcal{B}}}\mathbf{U}.
    \label{eq:LTI}
\end{equation}
Thus, the quasilinear form \eqref{eq:lin_calB_eq_finite} is equivalent to an LTI representation.
\end{proof}
A necessary and sufficient condition for the controllability of a linear system defined by the pair $(A, B)$ is that the augmented matrix $\left[A - \lambda I \quad B\right]$ has full row rank at every eigenvalue $\lambda$ of $A$ (cf. \cite{chen1984linear}, Th. 6.1).     By construction, the system matrix $\mathbf{A}$ in \eqref{eq:LTI} and \eqref{eq:lin_calB_eq_finite} has a repeated eigenvalue at $\lambda=0$. We summarize the controllability of the lifted LTI and quasilinear model in Proposition \ref{prop:controllability}.

\begin{proposition}
\label{prop:controllability}
The lifted LTI system \eqref{eq:LTI} is controllable. Moreover, the controllability of \eqref{eq:LTI} ensures that the quasilinear system \eqref{eq:lin_calB_eq_finite} is controllable.
\end{proposition}

\begin{proof}
The augmented matrix \([\mathbf{A} \ \overline{\boldsymbol{\mathcal{B}}}]\) achieves full row rank through the synergistic structure of \(\mathbf{A}\) and \(\overline{\boldsymbol{\mathcal{B}}}\):  
The block-diagonal matrix \(\mathbf{A} = \operatorname{blkdiag}(\mathbf{A}_p, \mathbf{A}_a)\) in \eqref{eq:define_calB} propagates states via identity sub-blocks (\(\mathbf{A}_1, \mathbf{A}_2, \mathbf{A}_3, \mathbf{A}_a\)) but introduces zero rows in its lower block rows, while \(\overline{\boldsymbol{\mathcal{B}}} = \operatorname{blkdiag}(\boldsymbol{\mathcal{B}}_1, \boldsymbol{\mathcal{B}}_2, \boldsymbol{\mathcal{B}}_3, \boldsymbol{\mathcal{B}}_4)\) injects complementary non-zero entries.  
Specifically, \(\boldsymbol{\mathcal{B}}_1, \boldsymbol{\mathcal{B}}_2, \boldsymbol{\mathcal{B}}_3\) populate the first, second, and third block rows of \(\mathbf{A}_p\), and \(\boldsymbol{\mathcal{B}}_4\) fills the zero rows of \(\mathbf{A}_a\) via identity sub-blocks.  
These identity terms act as pivots, eliminating rank deficiencies in \(\mathbf{A}\), ensuring linear independence of all rows in \([\mathbf{A} \ \overline{\boldsymbol{\mathcal{B}}}]\). This ensures the controllability of \eqref{eq:LTI}.  

The lifted input \(\mathbf{U} = \tilde{\boldsymbol{\mathcal{B}}} \widetilde{\mathbf{u}}\) is lossless and bijective because \(\tilde{\boldsymbol{\mathcal{B}}}\) retains full column rank. This is ensured by two properties:  
(i) the first column of \(\tilde{\boldsymbol{\mathcal{B}}}\) contains a single non-zero entry \(1/m\) from \(\bar{\mathbf{B}}_{M+1}\), isolating the thrust input dynamics, and  
(ii) the block \(\mathbf{B}_{3M+2} = (\mathbf{I}_3 \otimes \mathbf{z}_1)\mathbf{G}^2\) preserves full rank. Here, \(\mathbf{G}^2 = \begin{bmatrix} \operatorname{vec}(\mathbf{j}_1^\times) & \operatorname{vec}(\mathbf{j}_2^\times) & \operatorname{vec}(\mathbf{j}_3^\times) \end{bmatrix}\) inherits full column rank from the full-rank structure of \(\hat{\mathbf{J}} = \begin{bmatrix} \mathbf{j}_1 & \mathbf{j}_2 & \mathbf{j}_3 \end{bmatrix}\), while the Kronecker product \(\mathbf{I}_3 \otimes \mathbf{z}_1\) (with $\mathbf{z}_1 = \mathbf{R}$) maintains rank due to \(\mathbf{R}\)'s invertibility.  Collectively, these properties guarantee \(\tilde{\boldsymbol{\mathcal{B}}}\) has linearly independent columns across all operating conditions, ensuring \(\mathbf{U}\) spans the input space without redundancy or loss of control authority. Therefore, the controllability of \eqref{eq:LTI} implies the controllability of \eqref{eq:lin_calB_eq_finite}.
\end{proof}

\vspace{-0.6em}

\section{Controller Synthesis}

Model Predictive Control (MPC) employs a nominal plant model and, at each sampling instant, solves a finite‐horizon optimal control problem (OCP) over a prediction horizon \(T_H\) discretized into \(N_H\) intervals. For the quadrotor system under study, adopting the quasilinear dynamics \eqref{eq:lin_calB_eq_finite} as the nominal model reduces the number of decision variables compared to the LTI formulation based on \eqref{eq:LTI}. However, the quasi‐linear structure of \eqref{eq:lin_calB_eq_finite} prevents the resulting MPC problem from being cast as a quadratic program (QP), while the LTI dynamics of \eqref{eq:LTI} readily admits such a formulation. In contrast, explicitly enforcing constraints on the lifted control vector \(\mathbf{U}\) in \eqref{eq:LTI} remains challenging. To reconcile these issues, we propose a QP‐solvable LMPC scheme that employs \eqref{eq:lin_calB_eq_finite} as the nominal model while imposing constraints on the control input and the states.

\subsection{Proposed KQ-LMPC Scheme}
Let $\mathbf{x}(t)$ denote the state of the quadrotor at time $t$, with the control input. The reference trajectory and control input are $\mathbf{x}_r(t)$ and $\mathbf{u}_r(t)$, respectively. 
The nominal state trajectory predicted using~\eqref{eq:lin_calB_eq_finite} over the MPC horizon at time~$t$ is denoted by $\bar{\mathbf{X}}(l \mid t)$ for $l \in [t,\, t + T_H]$, where the horizon is discretized with time step $\delta = T_H / N_H$ and integrated using a 4th-order Runge--Kutta scheme.
With a reference trajectory $\mathbf{x}_r(t)$, the tracking error is $\hat{e}(t) = \bar{\mathbf{X}}(t) - \mathbf{X}_r(t)$, where $\mathbf{X}_r(t) = \boldsymbol{\varphi}(\mathbf{x}_r(t))$, and $\hat{e}(l|t) = \bar{\mathbf{X}}(l|t) - \mathbf{X}_r(l|t)$ for $l \in [t,t+T_H]$. Given control sampling interval $dt$, the optimal control problem (OCP) is formulated as: 
\begin{subequations}\label{eq:lmpc}
\begin{align}
       \! \! \! \mathcal{P}  & : V_{T_H}\big( \hat{e}(t)\big)  = \underset{\mathbf{u}(l|t)}{\text{  min   }} \mathcal{J}\big(\hat{e}(l|t), \mathbf{u}(l|t)\big)\\
            \label{eq:ocp_quasi}
       \! \! \! \!  \!  \text{s.t. : } 
        & \bar{\mathbf{X}}(t|t) \!=\! \boldsymbol{\varphi}\big( \mathbf{x}(t) \big), \ \!
         \widetilde{\mathbf{u}}(l|t) \!=\! \mathscr{U}\big( \boldsymbol{\varphi}^{\!-1}\big( \mathbf{\bar{X}}^{*}_{-1}\big),\mathbf{u}(l|t) \big), \\
       \! \! \! &  \dot{\bar{\mathbf{X}}}(l|t) = \mathbf{A\bar{X}}(l|t) + \boldsymbol{\mathcal{B}}\big(\mathbf{\bar{X}}^{*}_{-1}\big)\widetilde{\mathbf{u}}(l|t), \label{eq:lpv} \\   
      \! \! \! &  \mathbf{u}(l|t) \in \mathcal{U},\ \boldsymbol{\varphi}^{-1}\big( \bar{\mathbf{X}}(l|t)\big) \in \boldsymbol{\mathcal{X}},
    \end{align}
\end{subequations}
with the objective function defined as 
\begin{equation}
   \! \!  \mathcal{J}\big(\hat{e}(l|t), \mathbf{u}(l|t)\big) = \int_{t}^{t+T_H} \! \! \! \! \! \! \! \big( ||\hat{e}(\tau|t)||_{\boldsymbol{\mathcal{Q}}}^2 + ||\hat{u}(\tau|t)||_{\boldsymbol{\mathcal{R}}}^2 \big) d\tau,
   \label{eq:mpc_obj}
\end{equation}
where $\mathbf{\bar{X}}^{*}_{-1}\! :=\!  \mathbf{\bar{X}}^{*}(l|t-dt)$ denotes the nominal state trajectory (hereafter referred to as \emph{previously predicted optimal trajectory} or  \emph{PPOT}) predicted at the previous sampling instant $t-dt$ using the optimal control sequence $\mathbf{u}^{*}(l\!\mid\!t-dt)$ obtained by solving the OCP \eqref{eq:lmpc}, input error $\hat{u}(l|t) = \mathbf{u}(l|t)-\mathbf{u}_r(l|t)$, $\boldsymbol{\mathcal{Q}} \succeq 0 $ is the weight matrix for the tracking error, and $\boldsymbol{\mathcal{R}} \succ 0$ is the weight matrix for the input error. Using PPOT $\mathbf{\bar{X}}^{*}_{-1}$, we reformulate the nominal system model \eqref{eq:lin_calB_eq_finite} as a linear parameter-varying (LPV) system  \eqref{eq:lpv}, thereby enabling QP solvability. The proposed LMPC scheme, termed \emph{KQ-LMPC}, is summarized in Algorithm~\ref{alg:lmpc}.

\noindent \emph{Constraints in KQ-LMPC Scheme:} The input constraint $\mathbf{u}\! \in \!\mathcal{U}$ is imposed as the inequality $\mathbf{L} \mathbf{u}\! \leq \!\mathbf{b}$, where $\mathbf{L}\! = \![-\mathbf{I}_4^\top \ \ \mathbf{I}^\top_4]^\top$ and $\mathbf{b} \!=\! [-\mathbf{u}_{lb}^\top, \ \mathbf{u}_{ub}^\top]^\top$. To ensure that constraints in the lifted space translate correctly to the original state space, we require $\boldsymbol{\varphi}^{-1}\big( \mathbf{X}(l|t)\big) \in \boldsymbol{\mathcal{X}}$, where $\boldsymbol{\varphi}^{-1}$ is nonlinear and $\mathbf{z}_1(l|t) \in SO(3)$ with initialization $\mathbf{z}_1(l|0) \in SO(3)$. In practice, the state constraints in the lifted space are imposed as  $\widebar{\mathbf{L}} \widehat{\mathbf{X}}(l|t) \leq \widebar{\mathbf{b}}(l|t)$, where $\widehat{\mathbf{X}}(l|t) = [\mathbf{p}_1^\top(l|t),\ \mathbf{y}_1^\top(l|t),\ \underline{\mathbf{z}}_2^\top(l|t)]^\top$, and $\widebar{\mathbf{L}} = [-\mathbf{I}_{15}^\top, \ \mathbf{I}_{15}^\top ]^\top$, with $\widebar{\mathbf{b}}(l|t) = [-\mathbf{p}_{1_{lb}}^\top, \ -\mathbf{y}_{1_{lb}}^\top,\ -\underline{\mathbf{z}}_{2_{lb}}^\top, \ \mathbf{p}_{1_{ub}}^\top, \ \mathbf{y}_{1_{ub}}^\top,\  \underline{\mathbf{z}}_{2_{ub}}^\top ]^\top$. Here, the bounds are specified from \eqref{eq:obs} using $\underline{\mathbf{z}}^*_1(l|t-dt)$ obtained from the \emph{PPOT}, together with $\mathbf{s}_{lb}$, $\mathbf{v}_{lb}$, $\boldsymbol{\omega}_{lb}$, $\mathbf{s}_{ub}$, $\mathbf{v}_{ub}$, and $\boldsymbol{\omega}_{ub}$. These constraints, together with the LPV dynamics \eqref{eq:lpv}, ensure that the resulting OCP $\mathcal{P}$ remains a QP.

\noindent \emph{Baseline Nonlinear MPC (NMPC)}: For comparison purposes, we consider an NMPC formulation similar to \cite{sun2022comparative} that employs the full nonlinear quadrotor model~\eqref{eq:quad_affine} while retaining the same horizon parameters and control sampling used in KQ-LMPC. The nominal state trajectory predicted at time $t$ is denoted by $\bar{\mathbf{x}}(l| t)$ for $l \in [t,\, t + T_H]$, and is computed using the nominal model~\eqref{eq:quad_affine} with 4th-order Runge--Kutta integration. We define the tracking error as $\hat{e}(t)\!=\!\mathbf{x}(t)-\mathbf{x}_r(t)$, and $\hat{e}(l|t)\!=\!\bar{\mathbf{x}}(l|t)-\mathbf{x}_r(l|t)$ for $l\in[t,t+T_H]$. Therefore, the OCP for the NMPC scheme is as follows:
\begin{subequations}\label{eq:nmpc}
\begin{align}
       \! \! \! \hat{\mathcal{P}} & : V_{T_H}\big( \hat{e}(t)\big)  = \underset{\mathbf{u}(l|t)}{\text{  min   }} \mathcal{J}\big(\hat{e}(l|t), \mathbf{u}(l|t)\big)\\
            \label{eq:ocp_nmpc}
        \text{s. t. : } 
        & \bar{\mathbf{x}}(t|t) = \mathbf{x}(t), \\
        &   \dot{\bar{\mathbf{x}}}(l|t) = \mathbf{F}\big( \bar{\mathbf{x}}(l|t), \mathbf{u}(l|t) \big)\\        
       &  \mathbf{u}(l|t) \in \mathcal{U},\ \bar{\mathbf{x}}(l|t) \in \boldsymbol{\mathcal{X}},
    \end{align}
\end{subequations}
where the objective function $\mathcal{J}(\cdot)$ follows the same form as in~\eqref{eq:mpc_obj}, except that the state-weight matrix $\boldsymbol{\mathcal{Q}}$ is dimensioned to match the nonlinear state vector~$\bar{\mathbf{x}}$.

\vspace{0.5em}

\begin{algorithm}
\small

\caption{KQ-LMPC Algorithm}\label{alg:lmpc}
\begin{algorithmic}[1] 
\STATE \textbf{Input:} $\boldsymbol{\mathcal{Q}}, \boldsymbol{\mathcal{R}}$, $t_f$, $\delta$, $dt$,$\mathbf{u}_{min}, \mathbf{u}_{max}$,\ $\mathbf{s}_{lb}$, $\mathbf{v}_{lb}$, $\boldsymbol{\omega}_{lb}$, $\mathbf{s}_{ub}$, $\mathbf{v}_{ub}$, $\boldsymbol{\omega}_{ub}$
\STATE $t \gets 0$
\WHILE{$0 \leq t \leq t_f$}
    \STATE Measure $\mathbf{x}(t)$
    \IF{$t==0$} 
        \STATE $\bar{\mathbf{X}}^*_{-1}\gets \mathbf{X}_r(l|t)$
    \ENDIF
    \STATE Solve $\mathcal{P}$ \eqref{eq:lmpc} for $\mathbf{u}^*(l|t)$, obtain $\bar{\mathbf{X}}^*(l|t)$
    \STATE Apply $\mathbf{u}(\tau) \gets \mathbf{u}^*(\tau|t), \ \tau \in [t,t+\delta)$
    \STATE $t \gets t+dt$
    \STATE Store $\bar{\mathbf{X}}^*_{-1} \gets \bar{\mathbf{X}}^*(l|t)$
\ENDWHILE
\end{algorithmic}
\end{algorithm}

\subsection{Closed-loop Stability and Robustness of KQ-LMPC}

Let $\boldsymbol{\epsilon}\!\in\!\mathbb{R}^{18}$ be a bounded disturbance in \eqref{eq:quad_affine}, inducing $\boldsymbol{\eta}\!\in\!\mathbb{R}^{\mathcal{N}}$ in \eqref{eq:lin_calB_eq_finite} as $\boldsymbol{\eta}_\epsilon=(\partial\boldsymbol{\varphi}/\partial\mathbf{x})\boldsymbol{\epsilon}$. Taking into account also the truncation error $\boldsymbol{\eta}_T$, we have $\boldsymbol{\eta}\!=\!\boldsymbol{\eta}_\epsilon+\boldsymbol{\eta}_T$ with $|\boldsymbol{\eta}|\!\leq\!\eta_b$.

\begin{lemma}\label{lem_st1}
Under Assumption 1  and the Lipschitz continuity of the lifted dynamics, the mismatch $\bar{\mathbf{e}}(l|t)=\bar{\mathbf{X}}(l|t)-\bar{\mathbf{X}}^*(l|t-dt)$ satisfies $|\bar{\mathbf{e}}(l|t)| \leq \gamma_e(dt)$, for $l\in[t,t+T_H]$, where $\gamma_e(\cdot)$ is class $\mathcal{K}$.
\end{lemma}
\begin{proof}
  To maintain readability, the proof is provided in \Cref{app:lem2}.
\end{proof}
\begin{lemma}\label{lem_st2}
The prediction error $\mathbf{e}_p(t)\!=\!\mathbf{X}(t)\!-\!\bar{\mathbf{X}}(t)$ over a finite horizon admits an ultimate bound given by
$
|\mathbf{e}_p(t)|\!\leq \! \gamma_{ep1}(dt)+\gamma_{ep2}(\eta_b),
$ where $\gamma_{ep1}(\cdot)$ and $\gamma_{ep2}(\cdot)$ are class $\mathcal{K}$.
\end{lemma}
\begin{proof}
For clarity of presentation, the detailed proof is deferred to \Cref{app:lem3}.
\end{proof}

\begin{theorem}\label{thm3}
Under Assumption 1 and the Lipschitz continuity of the lifted dynamics, the KQ-LMPC scheme in closed loop is \emph{input-to-state practically stable} (ISPS) with respect to disturbance $\boldsymbol{\eta}$, with $\mathbf{X}$ converging to a bounded set of size determined by $\eta_b$ and $dt$.
\end{theorem}
\begin{proof}
The proof is omitted here for brevity and can be found in \Cref{app:thm3}.
\end{proof}

\begin{remark}
Although the current KQ-LMPC formulation achieves reliable closed-loop performance through a carefully structured constraint design and consistent empirical feasibility, we acknowledge that establishing formal recursive feasibility guarantees remains an open theoretical direction. In this work, we have not incorporated terminal ingredients (such as terminal sets or costs) because the computational efficiency of KQ-LMPC enables the use of sufficiently long prediction horizons in real time, which has been shown in practice to preserve feasibility without resorting to additional terminal constraints. The ISPS framework ensures that the closed-loop system states remain bounded within a practical region determined by the magnitude of the disturbance $\boldsymbol{\eta}$.
\end{remark}

\begin{remark}
While the proposed KQ-LMPC exhibits reliable closed-loop performance across all experimental evaluations, we acknowledge the possibility of rare infeasibility events. To enhance practical safety, we suggest incorporating a fallback LQR controller (see \Cref{sub:lqr}) that temporarily stabilizes the system whenever the KQ-LMPC optimization becomes infeasible. This safeguard acts as a feasibility recovery layer, ensuring continuity of control without modifying the nominal MPC behavior during standard operation. The KQ-LMPC algorithm with LQR fallback mechanism is presented in \Cref{alg:lmpc_lqr}.
\end{remark}
\vspace{-0.6em}
\section{Numerical and Experimental Results}

This section presents numerical simulations and experimental validations for the proposed Koopman linearization procedure and KQ-LMPC scheme. A comparative analysis with the baseline NMPC scheme is also provided to demonstrate the effectiveness of our approach. The custom-built quadrotor considered in this study (Figure~\ref{fig:quad_model}(b)) integrates an NVIDIA Jetson NX onboard computer, a Cube Orange Plus flight controller running PX4 firmware, and a Vicon Vantage motion capture system for state estimation. 
The quadrotor parameters are: $m  \!=\! 0.904$ kg and $\mathbf{J}\!=\!\text{diag}([0.00235,\, 0.00263,\, 0.00319])\ \text{kg}\cdot\text{m}^2$.
The reference state trajectory is denoted by $\mathbf{x}_r(t) \in \boldsymbol{\mathcal{X}}$. Here, input and rotational references are constructed geometrically as in \cite{lee_geom}.  Based on translational references, with $t\! \in \! [0,t_f]$, we consider four tasks for comparison of the proposed KQ-LMPC and baseline NMPC methods: 
(1) \emph{\textbf{Task 1} (Follow a Vertical Line and Hover): } $\mathbf{s}_r(t)$ is a fifth-order polynomial satisfying $\mathbf{s}_r(0) = \mathbf{s}_{\text{oc}},\ \mathbf{s}_r(t_f) =  \mathbf{s}_{\text{oc}}+ 2 \overrightarrow{e_3},\ \text{and } \mathbf{v}_r(t_f) = \mathbf{0}_{3\times 1} \ $;  (2) \emph{\textbf{Task 2} (Vertical Helix): } $\mathbf{s}_r(t) = [ \cos(0.4 t), \  \sin(0.4 t),\ z_0+ t/80]^\top$; (3) \emph{\textbf{Task 3} (Lemniscate): }  $\mathbf{s}_r(t) = [ \sin(0.8 t), \  \sin(0.8 t) \cos(0.8 t),\ z_0]^\top$; and (4) \emph{\textbf{Task 4} (Knot): } $\mathbf{s}_r(t) = [ 0.8+0.6\cos(1.2t)\cdot\cos(0.8t), \  0.8+0.6\cos(1.2t)\cdot\sin(0.8t),\ z_0+ 0.6\sin(1.2t)]^\top$.
Here, $\mathbf{s}_{\text{oc}}\!=\![x_0, y_0, z_0]^\top$ in $\mathcal{I}_{\text{ref}}$ marks the transition point where the onboard low-level controller relinquishes control and the MPC schemes take command. The tasks are designed to span a range of dynamic performance: Tasks 1 and 2 represent low to moderate agility, while Tasks 3 and 4 introduce reasonably agile maneuvering demands.

For the KQ-LMPC scheme, we set \(\mc{N} \!=\! 45\) ($M\!=\!3$, $N\!=\!2$) and the state weighting matrix as \(\boldsymbol{\mathcal{Q}} = \operatorname{blkdiag}(10^3 \mathbf{I}_3,\ 500 \mathbf{I}_3,\ \mathbf{0}_3,\ 500 \mathbf{I}_6,\ \mathbf{0}_3,\ 600 \mathbf{I}_9,\ 200 \mathbf{I}_9)\). For the NMPC scheme, we use a uniform state weighting of \(\boldsymbol{\mathcal{Q}} = 10^3 \mathbf{I}_{18}\). For both schemes, we set \(\boldsymbol{\mathcal{R}} = \operatorname{diag}(10^{-3},\ 10^{-4}\mathbbm{1}_2^\top)\), with a control update interval $dt=0.01$ s and an MPC discretization step of \(\delta = 0.2~\text{s}\). 
The input bounds derived from the quadrotor parameters are $\mathbf{u}_{ub} \!=\! [30.56,\ \boldsymbol{\tau}_{ub}]^\top$ and $\mathbf{u}_{lb} \! =\! [0,\ -\boldsymbol{\tau}_{ub}]^\top$, where $\boldsymbol{\tau}_{ub} \!=\! [0.764\mathbbm{1}_2^\top,\ 0.0378]$. The state bounds are given as $\mathbf{s}_{ub} \!=\! 2\mathbbm{1}_2$, $ \mathbf{s}_{lb}  \!=\! -\mathbf{s}_{ub}, \ \mathbf{v}_{ub}  \!=\! 5\mathbbm{1}_2,\ \mathbf{v}_{lb}  \!=\! -\mathbf{v}_{ub},\ \boldsymbol{\omega}_{ub}  \!=\! 0.7\mathbbm{1}_2, \  \boldsymbol{\omega}_{lb}  \!=\! -\boldsymbol{\omega}_{ub}$. All quadratic sub-problems in the proposed KQ-LMPC controller are solved with the high-performance interior-point solver \emph{HPIPM}, whereas the nonlinear sub-problems of the baseline NMPC are handled by the \emph{SQP} method from the \emph{acados} toolchain \cite{verschueren2022acados} through its Python interface. The tracking performance of the KQ-LMPC and NMPC schemes is compared over a time sequence $\{t_k\}_{k=0}^{\mathfrak{N}}$, where $t \in [0, t_f]$ with a uniform plant sampling interval $\Delta t = 5 \text{ ms}$. The performance metric is the root mean squared error of tacking defined as
$
    \mathcal{E}_s = \sqrt{(1/\mathfrak{N}){\scriptstyle\sum_{k=1}^{\mathfrak{N}}}|\mathbf{s}(t_k)-\mathbf{s}_r(t_k)}|^2.
$


\subsection{Approximation Error in Lifted Formulation}

\begin{figure}[!htpb]
\centering
\includegraphics[scale=0.28]{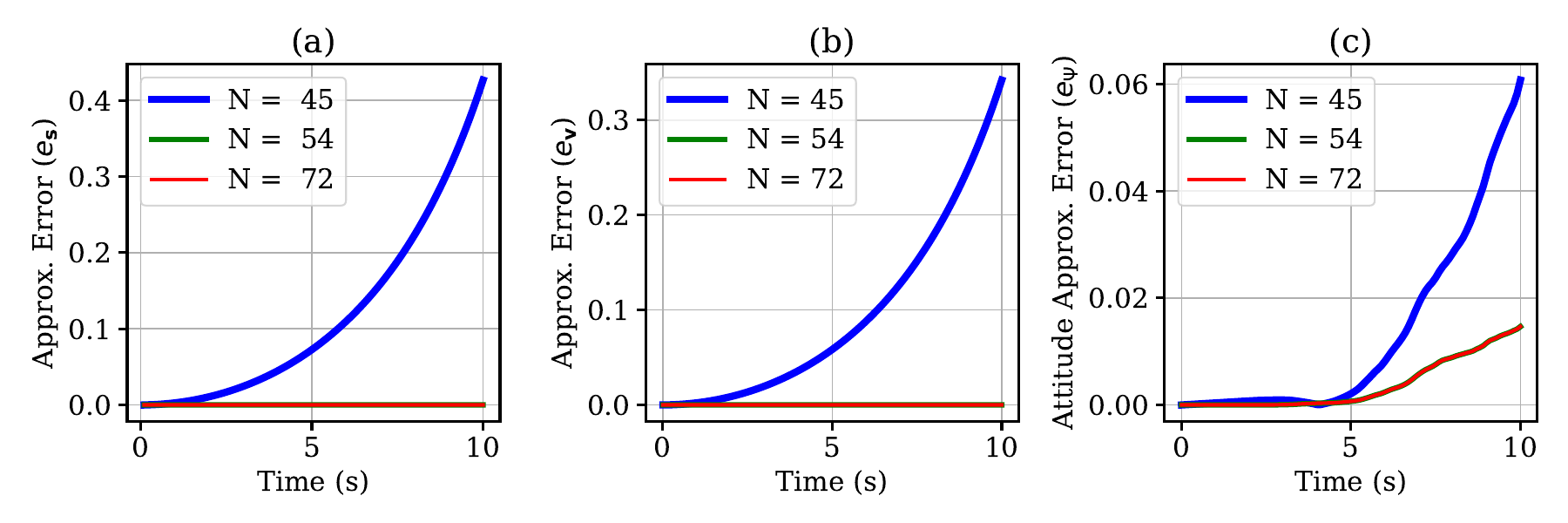}
\caption{Approximation errors for a period of 10 sec with different lifted-state dimension.} 
\label{fig:appxerr}
\end{figure}

The accuracy of the lifted quasilinear representation \eqref{eq:lin_calB_eq_finite}, with trajectories denoted by $(\cdot)_{ql}$, is evaluated by comparing against the exact non-linear model \eqref{eq:quad_compact}, with trajectories indicated by $(\cdot)_{nl}$. The \emph{relative approximation error} is defined as
$ e_{\phi}(t) \!=\! 
    | \phi_{ql}(t) - \phi_{nl}(t) |/
         | \phi_{nl}(t) |,
$
where $\phi \in \{\mathbf{s},\,\mathbf{v}\}$.  The \emph{attitude approximation error} is given by
$ e_{\psi} \!=\! \tfrac{1}{2}\,\big(\mathbf{I}_3 - \mathbf{R}_{ql}^\top(t)\, \mathbf{R}_{nl}(t)\big).
$ Simulations start from $\mathbf{s}(0)\!=\!\mathbf{0}_{3\times1}$, $\mathbf{v}(0)\!=\!0.1[1,1,1]^\top$, $\mathbf{R}(0)\!=\!\mathbf{I}_3$, $\boldsymbol{\omega}(0)\!=\!0.05[1,1,1]^\top$, with control input $\widetilde{\mathbf{u}}(t)\!=\!\boldsymbol{\kappa}(t)\sin(0.1t)$, where $\boldsymbol{\kappa}(t)\in\mathbb{R}^4$ is drawn uniformly at random  from $[-0.005,0.005]$. We observe that as the dimension of the lifted space increases, the approximation error decreases. However, irrespective of the dimension, the error accumulates over time due to the integration of truncated terms. The approximation errors for a period of 10 $s$ is depicted in \Cref{fig:appxerr}. At \( t = 5  \text{s} \), the approximation errors for \( \mathcal{N} = 45, 54, \) and \(72 \) are \( \{ e_{\mathbf{s}} \! =\!0.073,\ e_{\mathbf{v}} \!=\! 0.058,\ e_{\psi} \!=\! 0.002 \} \), \( \{ e_{\mathbf{s}} \!=\! 2.1\cdot10^{-5}, \ e_{\mathbf{v}} \!=\! 10^{-5},\ e_{\psi} \!=\! 6.8\cdot10^{-4} \} \), and \( \{ e_{\mathbf{s}} \!=\! 10^{-6}, \ e_{\mathbf{v}} \!=\! 10^{-6},\ e_{\psi} \!=\! 6.7\cdot10^{-4} \} \), respectively. Compared to \cite{zinage} with $\mathcal{N}\!=\!61$, the proposed quasilinear formulation with $\mathcal{N}\!=\!54$ reduces position and velocity errors by approximately three orders of magnitude, with only a minor (< one order) increase in attitude error. Given that the MPC horizon $T_H$ is less than $5 \ \text{s}$, we adopt $\mathcal{N}\! =\! 45$ for the nominal model in the MPC formulation. 

\subsection{Numerical Simulation Results}

Using numerical simulations, we benchmark the proposed KQ-LMPC against the baseline NMPC, evaluating both computational load and tracking accuracy across various tasks and prediction horizons. For each task-horizon combination, two $10,\text{s}$ closed-loop simulations of the plant model \eqref{eq:quad_dynamics1} (with uniform random process noise in each component of the state drawn from $[\scriptsize{-10^{-3},10^{-3}}])$ were conducted. The solver runtime was recorded at each control interval. The mean computation time ($\mu_t$) provides a measure of average algorithmic efficiency, while the worst-case computation time ($t_w$) is the critical metric for guaranteeing real-time feasibility. As shown in \Cref{tab1}, KQ-LMPC consistently achieved lower $\mu_t$ than NMPC across all cases. For instance, at $T_H = 2.0$ s, $\mu_t$ for KQ-LMPC was 0.78–0.87 ms, compared to 1.69–3.23 ms for NMPC. Moreover, $\mu_t$ for KQ-LMPC was nearly insensitive across tasks due to its convex QP formulation, whereas NMPC’s non-convex optimization produced higher and more variable runtimes. Notably, $\mu_t$ for KQ-LMPC remained nearly insensitive to task agility owing to its convex QP formulation, while NMPC’s non-convex optimization led to higher and more variable computation times. Importantly, $t_w$ for KQ-LMPC consistently remained below 50\% of the control sampling time $dt$, ensuring real-time feasibility, whereas NMPC in several cases exceeded $dt$ or reached more than 50\% of $dt$. In terms of tracking accuracy, KQ-LMPC matched or slightly outperformed NMPC for the less agile tasks 1–2, as seen in RMSE values ($\mathcal{E}_s$). For the more agile tasks 3–4, NMPC held a modest advantage with $\mathcal{E}_s \leq 12$ cm, while KQ-LMPC maintained respectable performance with $\mathcal{E}_s \leq 18$ cm. Overall, KQ-LMPC offers a favorable balance of computational efficiency and tracking accuracy, delivering consistent runtime behavior suited for embedded platforms while maintaining competitive control performance. Finally, we select $T_H = 2.0$ s for experimental implementation, as it consistently yielded low $\mathcal{E}_s$ across tasks.

\begin{table*}[t]
\centering

\caption{Numerical Simulation (performed using Python 3.10 with Ubuntu 22.04 and Ryzen 3 PRO CPU): mean computation time ($\mu_t$), worst-case computation time ($t_w$), and tracking RMSE ($\mathcal{E}_s$) for KQ-LMPC (proposed) and NMPC (baseline)}.

\resizebox{\textwidth}{!}{%
\tiny
\begin{tabular}{l|l|cccc|cccc|cccc|cccc}
\hline
 & & \multicolumn{4}{c|}{\textbf{$T_H = 0.8$ s}}
   & \multicolumn{4}{c|}{\textbf{$T_H = 1.4$ s}}
   & \multicolumn{4}{c|}{\textbf{$T_H = 2.0$ s}}
   & \multicolumn{4}{c}{\textbf{$T_H = 2.8$ s}} \\
\hline
 & \textbf{Task} 
 & \textbf{1} & \textbf{2} & \textbf{3} & \textbf{4}
 & \textbf{1} & \textbf{2} & \textbf{3} & \textbf{4}
 & \textbf{1} & \textbf{2} & \textbf{3} & \textbf{4}
 & \textbf{1} & \textbf{2} & \textbf{3} & \textbf{4} \\
\hline
\multirow{3}{*}{KQ-LMPC} 
 & $\mu_t$(ms) & 0.32 & 0.32 & 0.33 & 0.34 
               & 0.47 & 0.47 & 0.51 & 0.50
               & 0.78 & 0.80 & 0.90 & 0.87
               & 1.04 & 1.07 & 1.31 & 1.23 \\
 & $t_{w}$(ms) & 0.79 & 0.76 & 1.00 & 0.98 
               & 1.28 & 1.26 & 1.48 & 1.26
               & 2.25 & 2.18 & 2.49 & 2.13
               & 2.63 & 2.85 & 5.00 & 3.32 \\
 & $\mathcal{E}_s$(m) 
               & 0.06 & 0.09 & 0.10 & 0.13
               & 0.05 & 0.06 & 0.14 & 0.18
               & 0.05 & 0.04 & 0.10 & 0.12
               & 0.05 & 0.05 & 0.14 & 0.15 \\
\hline
\multirow{3}{*}{NMPC} 
 & $\mu_t$(ms) & 0.86 & 0.97 & 1.18 & 1.46 
               & 1.14 & 1.20 & 1.68 & 2.05
               & 1.69 & 1.75 & 2.70 & 3.24
               & 1.96 & 2.13 & 3.35 & 4.15 \\
 & $t_{w}$(ms) & 2.38 & 2.20 & 2.46 & 3.88 
               & 3.07 & 4.12 & 5.25 & 6.48
               & 4.48 & 4.53 & 8.52 & 9.47
               & 4.66 & 6.60 & 10.68 & 11.78 \\
 & $\mathcal{E}_s$(m) 
               & 0.05 & 0.07 & 0.09 & 0.09
               & 0.06 & 0.06 & 0.10 & 0.12
               & 0.04 & 0.06 & 0.06 & 0.07
               & 0.04 & 0.05 & 0.08 & 0.09 \\
\hline
\end{tabular}}  \vspace{-2em}
\label{tab1}
\end{table*}

\subsection{Experimental Evaluation}

\begin{figure*}[htpb!]
\centering
\subfloat[\scriptsize  $0.156$(KQ-LMPC), $0.155$(NMPC)]{\includegraphics[trim=0cm 5cm 0cm 5cm, clip=true, width=0.25\textwidth]{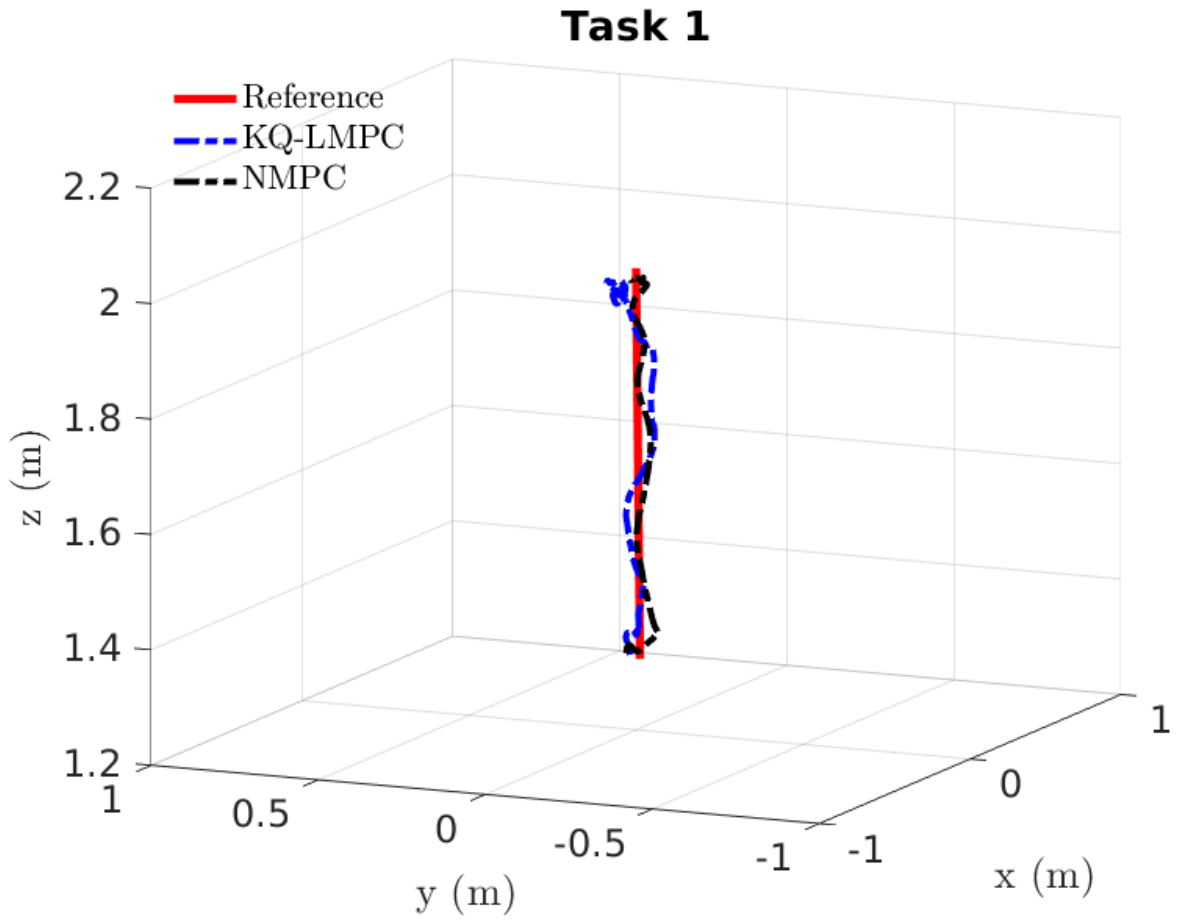}}
\subfloat[\scriptsize $0.186$(KQ-LMPC), $0.160$(NMPC)]{\includegraphics[trim=0cm 5cm 0cm 5cm, clip=true, width=0.25\textwidth]{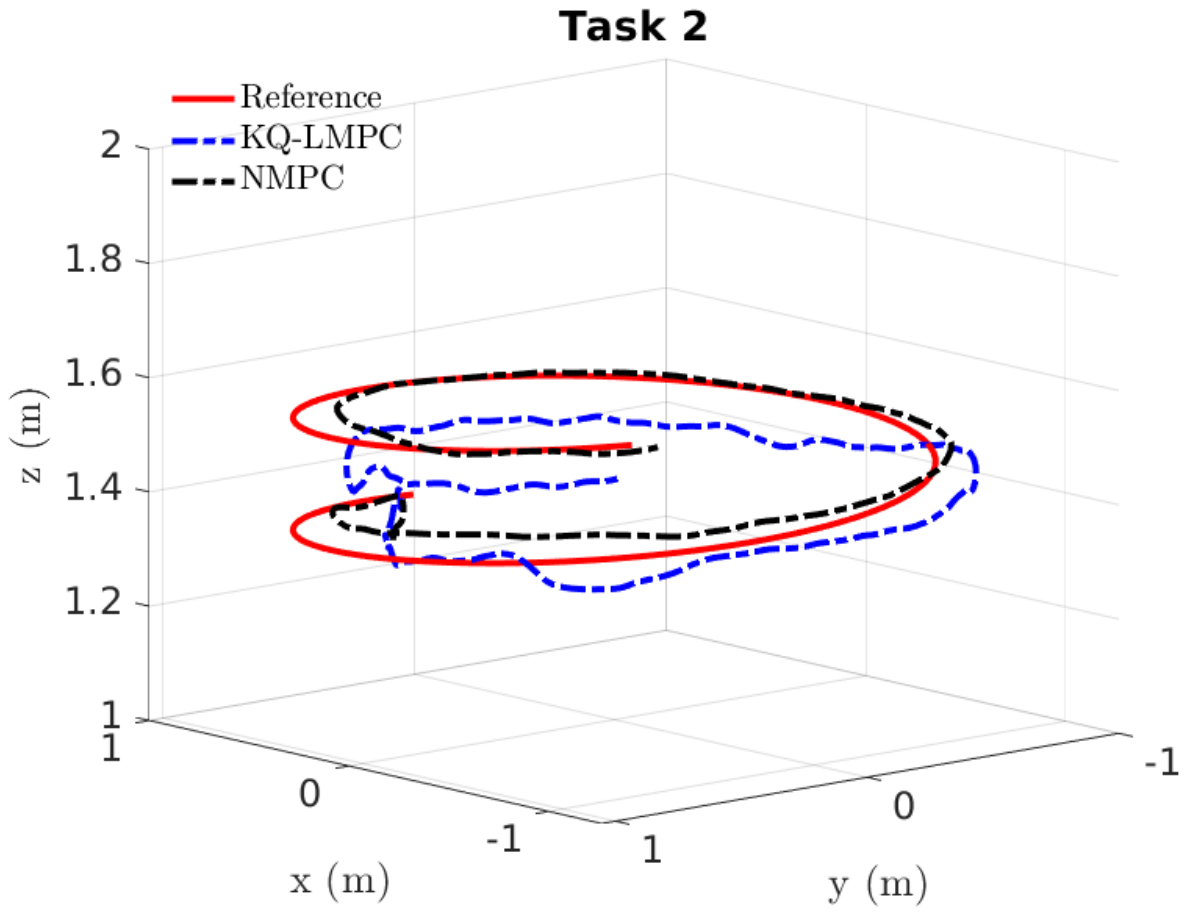}}
\subfloat[\scriptsize  $0.292$(KQ-LMPC), $0.258$(NMPC)]{\includegraphics[trim=0cm 5cm 0cm 5cm, clip=true, width=0.25\textwidth]{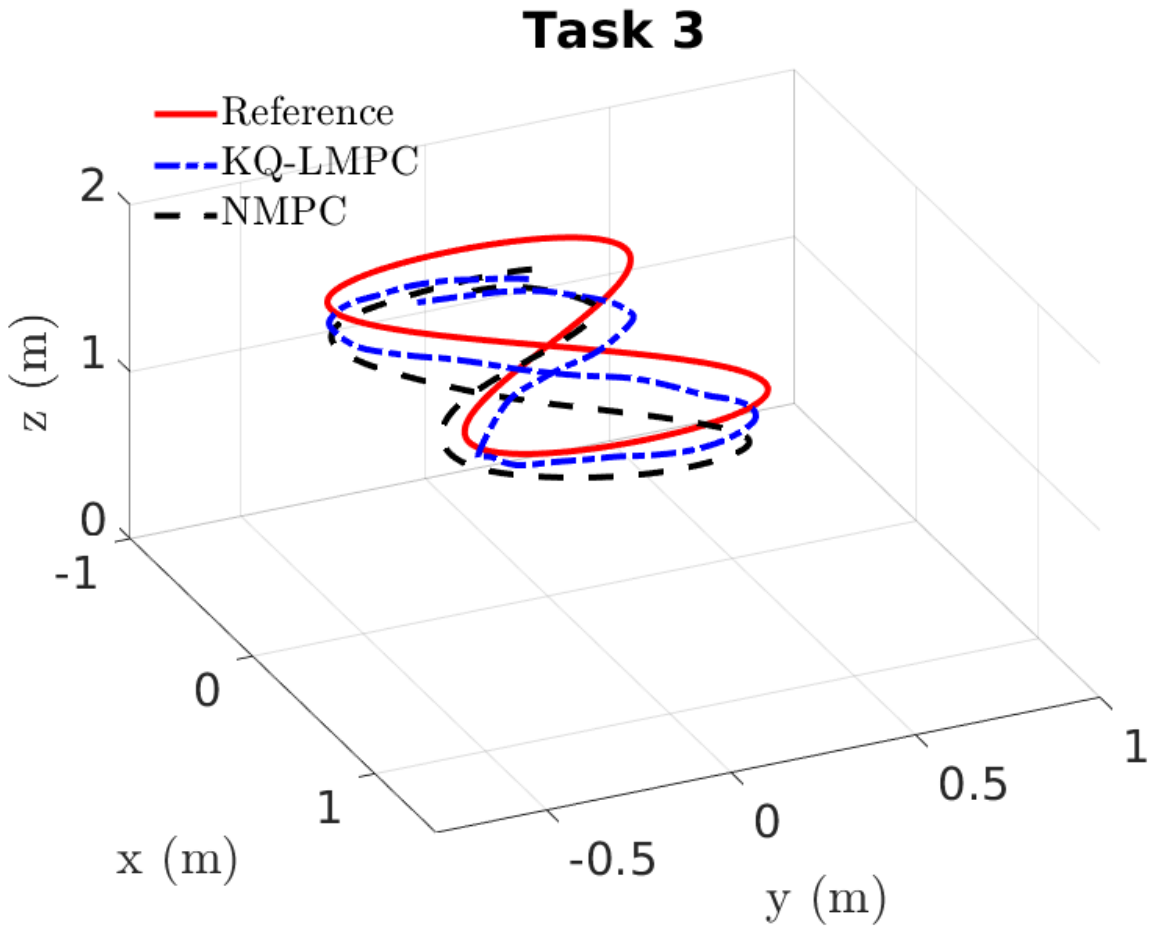}}
\subfloat[\scriptsize $0.298$(KQ-LMPC), $0.261$(NMPC)]{\includegraphics[trim=0cm 5cm 0cm 5cm, clip=true, width=0.25\textwidth]{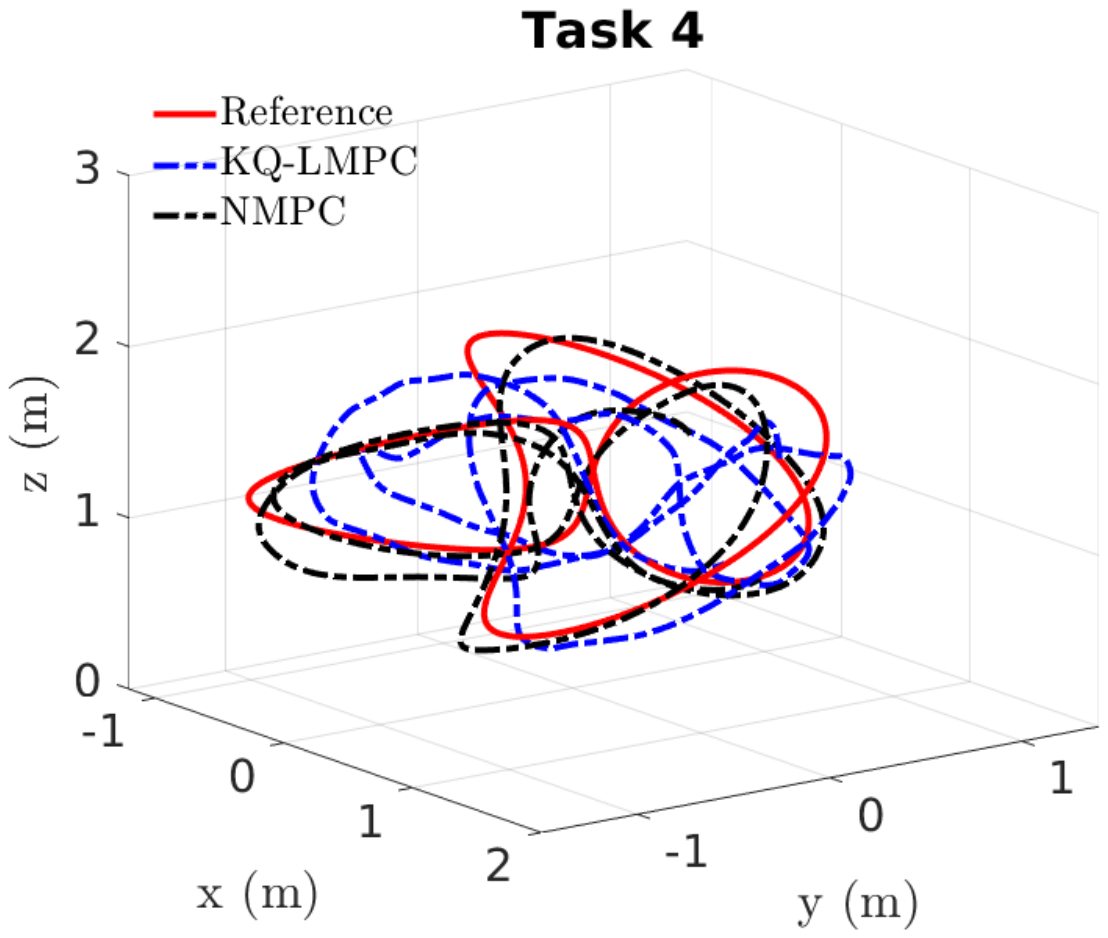}}
\caption{Experimental validation: trajectory tracking performance of KQ-LMPC (proposed) and NMPC (baseline). The values reported above are the tracking RMSEs ($\mathcal{E}_s$) in meters.}
\label{fig:tasks} 
\end{figure*}

To evaluate the real-world performance of the proposed KQ-LMPC scheme, we conducted experimental tests against the baseline NMPC across four benchmark tasks, with tracking results shown in \Cref{fig:tasks}. For the low-agility task (Task 1), KQ-LMPC achieved nearly identical accuracy to NMPC, with only a 0.1 cm difference in tracking RMSE ($\mathcal{E}_s$). As trajectory agility increased, $\mathcal{E}_s$ rose for both controllers, yet KQ-LMPC consistently delivered competitive performance. For the moderately agile task (Task 2), the difference in $\mathcal{E}_s$ was modest at 2.6 cm, while for the more agile trajectories (Tasks 3–4), the gap remained small, within a sub-4.0 cm margin (3.4 cm and 3.7 cm, respectively). These results demonstrate that KQ-LMPC maintains reliable tracking accuracy compared to NMPC while offering the critical advantage of substantially lower computational burden, ensuring real-time feasibility. Notably, for the reasonably agile Tasks 3 and 4, both controllers achieved competitive trajectory tracking, with $\mathcal{E}_s$ in the range of 20–30 cm. Further improvements may be realized through refined tuning of the MPC weighting matrices $\boldsymbol{\mathcal{Q}}$ and $\boldsymbol{\mathcal{R}}$, better sensing, and a robust extension of KQ-LMPC.

\vspace{-0.6em}
\section{Conclusion}\label{sec:conclusion}
This letter introduced an analytically derived Koopman-based linear embedding for quadrotor dynamics on SE(3), enabling an LPV representation that preserves controllability while admitting a tractable finite-dimensional approximation. Building on this lifted model, we developed the KQ-LMPC framework, which enforces system constraints through a convex QP and achieves real-time computational feasibility. Numerical simulations and experimental tests on reasonably agile tasks demonstrate that KQ-LMPC delivers tracking performance comparable to NMPC while significantly reducing computational burden, thus narrowing the performance-complexity gap in quadrotor MPC. Future work will extend this approach to more general rigid-body systems, such as fixed-wing aircraft. 
\vspace{-0.6em}

\section{Appendix}

\subsection{Proofs on Closed-Loop Stability}

\subsubsection{Proof of \Cref{lem_st1}} \label{app:lem2}
Consider the ultimate bounds $|\widetilde{\mathbf{u}}| \leq \widetilde{u}_b$, $|\mathbf{X}| \leq X_b$, and $\| \boldsymbol{\mathcal{B}}(\cdot) \| \leq B_b$.
The predicted state $\bar{\mathbf{X}}(l|t)$ satisfies:
\begin{equation}
\dot{\bar{\mathbf{X}}} = \mathbf{A} \bar{\mathbf{X}} + \boldsymbol{\mathcal{B}}(\bar{\mathbf{X}}^*_{-1}) \tilde{\mathbf{u}},
  \label{eq:lem1}
\end{equation}
while the PPOT $\bar{\mathbf{X}}^*_{-1} \equiv \bar{\mathbf{X}}^*(l|t-dt)$ (predicted at $t-dt$) satisfies:
\begin{equation}
\dot{\bar{\mathbf{X}}}^*_{-1} = \mathbf{A} \bar{\mathbf{X}}^*_{-1} + \boldsymbol{\mathcal{B}}(\bar{\mathbf{X}}^*_{-2}) \tilde{\mathbf{u}}^*_{-1},
 \label{eq:lem2}
\end{equation}
where $\bar{\mathbf{X}}^*_{-2} \equiv \bar{\mathbf{X}}^*(l|t-2dt)$ (PPOT for time step $t-dt$ evaluated at $t-2dt$) and $\tilde{\mathbf{u}}^*_{-1} = \tilde{\mathbf{u}}^*(l|t-dt)$. We define the mismatch between prediction and the PPOT as
\begin{equation}
\bar{\mathbf{e}}(l|t) = \bar{\mathbf{X}}(l|t) - \bar{\mathbf{X}}^*_{-1},
\end{equation}
with $\bar{\mathbf{e}}(t|t) = \mathbf{X}(t|t)-\bar{\mathbf{X}}^*(t|t-dt)$. Now, $\bar{\mathbf{X}}^*(t|t-dt)$ can be expressed as
\begin{equation}
\begin{aligned}
   \!\!\! \bar{\mathbf{X}}^*(t|t-dt) & = \mathbf{X}(t-dt)+\int_{t-dt}^t \bigg( \mathbf{A} \bar{\mathbf{X}}^*(\tau|t-dt)+\\
    & \quad \quad \quad  \boldsymbol{\mathcal{B}}\big(\bar{\mathbf{X}}^*(\tau|t-2dt)\big)\widetilde{\mathbf{u}}^*(\tau|t-dt)\bigg)d\tau \\
    & = \mathbf{X}(t-dt)+ \int_{t-dt}^t \frac{\partial}{\partial\tau} \dot{\bar{\vect{X}}}^*(\tau|t-dt)d\tau.
\end{aligned}
\end{equation}
Therefore,
\begin{equation}
\begin{aligned}
    \bar{\mathbf{e}}(t|t) & \leq |\mathbf{X}(t)-\mathbf{X}(t-dt)| + (\|\mathbf{A}\|X_b + B_b\widetilde{u}_b)dt, \\
    & \leq L_c dt,
\end{aligned}
\end{equation}
where $L_c = L_X+\|\mathbf{A}\|X_b + B_b\widetilde{u}_b$ with the Lipschitzness of the state vector as $|\mathbf{X}(t)-\mathbf{X}(t-dt)| \leq L_X\ dt$. Subtracting \eqref{eq:lem2} from \eqref{eq:lem1} yields:
\begin{equation}
\begin{aligned}
\dot{\bar{\mathbf{e}}} =& \mathbf{A} \bar{\mathbf{e}} + \boldsymbol{\mathcal{B}}(\bar{\mathbf{X}}^*_{-1}) \tilde{\mathbf{u}} - \boldsymbol{\mathcal{B}}(\bar{\mathbf{X}}^*_{-2}) \tilde{\mathbf{u}}^*_{-1} \\
=& \mathbf{A} \bar{\mathbf{e}} + \boldsymbol{\mathcal{B}}(\bar{\mathbf{X}}^*_{-1}) \Delta \tilde{\mathbf{u}} + \left[ \boldsymbol{\mathcal{B}}(\bar{\mathbf{X}}^*_{-1}) - \boldsymbol{\mathcal{B}}(\bar{\mathbf{X}}^*_{-2}) \right] \tilde{\mathbf{u}}^*_{-1},
\end{aligned}
\end{equation}
where $\Delta \tilde{\mathbf{u}} = \tilde{\mathbf{u}} - \tilde{\mathbf{u}}^*_{-1}$.
Let $\Delta \mathbf{X}^*_{\mathrm{prev}} = \bar{\mathbf{X}}^*_{-1} - \bar{\mathbf{X}}^*_{-2}$. Taking norm above gives the inequality,
\begin{equation}
|\dot{\bar{\mathbf{e}}}| \leq \|\mathbf{A}\| \, |\bar{\mathbf{e}}| + B_b |\Delta \tilde{\mathbf{u}}| + L_{\mathcal{B}} \, |\Delta \mathbf{X}^*_{\mathrm{prev}}| \, \tilde{u}_b,
\label{eq:lem1_ebardot}
\end{equation}
where the Lipschitzness properties  $|\Delta \tilde{\mathbf{u}}| \leq L_u dt$, and $\|\boldsymbol{\mathcal{B}}(\bar{\mathbf{X}}^*_{-1}) - \boldsymbol{\mathcal{B}}(\bar{\mathbf{X}}^*_{-2})\| \leq L_\mathcal{B}|\Delta \mathbf{X}^*_{\mathrm{prev}}|$ are valid. 
Now,
\begin{equation}
\begin{aligned}
    \Delta \mathbf{X}^*_{\mathrm{prev}} &=  \bar{\mathbf{X}}^*_{-1} - \bar{\mathbf{X}}^*_{-2} \\
    & = \int_{t-2dt}^{t-dt}\frac{\partial  }{\partial \tau} \mathbf{\bar{X}}^*(l|\tau) d\tau \\
    & = \int_{t-2dt}^{t-dt} \bigg( \mathbf{A}\mathbf{\bar{X}}^*(l|\tau) + \boldsymbol{\mathcal{B}}(\mathbf{\bar{X}}^*_{-1})\widetilde{\mathbf{u}}^*_{-1}. \bigg)d\tau.
\end{aligned}
\end{equation}
Therefore,
\begin{equation}
\begin{aligned}
    |\Delta \mathbf{X}^*_{\mathrm{prev}}| &\leq (\|\mathbf{A}\|X_b + B_b\widetilde{u}_b)dt\\
    & \leq (A_bX_bB_b+\widetilde{u}_b)dt,
\end{aligned}
\end{equation}
where $\|\mathbf{A}\|=A_b$. It follows from \eqref{eq:lem1_ebardot} that
\begin{equation}
\begin{aligned}
|\dot{\bar{\mathbf{e}}}| & \leq \|\mathbf{A}\| \, |\bar{\mathbf{e}}| + L_uB_bdt+L_\mathcal{B}\widetilde{u}_b(A_bX_bB_b+\widetilde{u}_b)dt \\
& \leq A_b  |\bar{\mathbf{e}}| + \big(\underbrace{L_uB_b+L_\mathcal{B}\widetilde{u}_b(A_bX_bB_b+\widetilde{u}_b)}_{C_{\bar{e}}}\big)dt.
\end{aligned}
\end{equation}
Using Gr$\ddot{\text{o}}$nwall's inequality, we obtain
\[
|\bar{\mathbf{e}}(l|t)| \leq \gamma_e(dt),
\]
where 
\[
\gamma_e(dt) = \bigg( L_c \,  \cdot e^{A_b T_H} + \frac{C_{\bar{e}}}{A_b} \left( e^{A_b T_H} - 1 \right)\bigg)dt
\]
is class $\mathcal{K}$.
This completes the proof.

\subsubsection{Proof of \Cref{lem_st2}}\label{app:lem3}
The actual system in the lifted space evolves as
\begin{equation}
    \Dot{\mathbf{X}} = \mathbf{A}\mathbf{X} + \boldsymbol{\mathcal{B}}(\mathbf{X})\widetilde{\mathbf{u}} + \boldsymbol{\eta}.
\end{equation}
The nominal MPC model evolves over a horizon as
\begin{equation}
    \dot{\bar{\mathbf{X}}}(l|t) = \mathbf{A\bar{X}}(l|t) + \boldsymbol{\mathcal{B}}\big(\mathbf{\bar{X}}^{*}_{-1}\big)\widetilde{\mathbf{u}}(l|t).
\end{equation}
The prediction error over an MPC horizon becomes
\begin{equation}
    \mathbf{e}_p(l|t) = \mathbf{X}(l|t)-\bar{\mathbf{X}}(l|t), 
\end{equation}
where $\mathbf{e}_p(t|t) = X(t|t)-\bar{X}(t|t) = \mathbf{0}$.
We can define the error dynamics as
\begin{equation}
    \dot{\mathbf{e}}_p \!=\! \mathbf{Ae}_p + \big(\boldsymbol{\mathcal{B}}(\mathbf{X})- \boldsymbol{\mathcal{B}}(\mathbf{\bar{X}})+\boldsymbol{\mathcal{B}}(\mathbf{\bar{X}})-\boldsymbol{\mathcal{B}}(\bar{\mathbf{X}}^{*}_{-1})\big)\widetilde{\mathbf{u}}+\boldsymbol{\eta}.
    \label{eq:err_dyn}
\end{equation}
Due to Lipschitzness of $\boldsymbol{\mathcal{B}}(\cdot)$, we have $\|\boldsymbol{\mathcal{B}}(\mathbf{X})- \boldsymbol{\mathcal{B}}(\mathbf{\bar{X}})\| \leq L_\mathcal{B}|\mathbf{e}_p|$, and $\|\boldsymbol{\mathcal{B}}(\mathbf{\bar{X}})-\boldsymbol{\mathcal{B}}(\bar{\mathbf{X}}^{*}_{-1})\| \leq L_\mathcal{B} |\bar{\mathbf{e}}|$, where $\bar{\mathbf{e}} = |\mathbf{\bar{X}}-\bar{\mathbf{X}}^{*}_{-1}|$. Therefore, from \eqref{eq:err_dyn},
\begin{equation}
\begin{aligned}
    |\dot{\mathbf{e}}_p| & \leq \|\mathbf{A}\||\mathbf{e}_p|+L_\mathcal{B}(|\mathbf{e}_p|+|\bar{\mathbf{e}}|)\widetilde{u}_b + \eta_b \\
    & \leq (A_b+L_\mathcal{B}\widetilde{u}_b)|\mathbf{e}_p|
    + L_\mathcal{B}\widetilde{u}_b \gamma_e(dt) + \eta_b \\
    & \leq M_e |\mathbf{e}_p| + L_\mathcal{B}\widetilde{u}_b \gamma_e(dt)  + \eta_b,
    \label{eq:epdot} 
\end{aligned}
\end{equation}
where $M_e  = A_b+L_\mathcal{B}\widetilde{u}_b$. Using Gr$\ddot{\text{o}}$nwall's inequality,
\begin{equation}
\begin{aligned}
    |\mathbf{e}_p(l|t)| & \leq \frac{L_\mathcal{B}\widetilde{u}_b \gamma_e(dt) + \eta_b}{M_e} (e^{M_e\ T_H}-1) \\
    &  \leq \ \underbrace{\frac{(e^{M_e\ T_H}-1)}{M_e} L_\mathcal{B}\widetilde{u}_b \gamma_e(dt)}_{\gamma_{ep_1}(dt)} \\
    & \quad \quad \quad \quad \quad + \underbrace{\frac{(e^{M_e\ T_H}-1)}{M_e} \eta_b}_{\gamma_{ep_2}(\eta_b)} \\
    & \leq \gamma_{ep_1}(dt)+\gamma_{ep_2}(\eta_b)
    \label{eq:ep_gr}
\end{aligned}
\end{equation}
where $\gamma_{ep_1}(\cdot)$  and $\gamma_{ep_2}(\cdot)$ are class $\mathcal{K}$. For brevity, consider $E_b = \gamma_{ep_1}(dt)$ and $C_b = \gamma_{ep_2}(\eta_b)$
This completes the proof.

\subsubsection{Proof of \Cref{thm3}} \label{app:thm3}

The actual system trajectory in terms of prediction error is 
\begin{equation}
    \mathbf{X}=\bar{\mathbf{X}}+\mathbf{e}_p.
\end{equation}
Consider the evolution of the actual system in the lifted space under closed-loop control $\widetilde{\mathbf{u}}^*$
\begin{equation}
    \Dot{\mathbf{X}} = \mathbf{A}\mathbf{X} + \boldsymbol{\mathcal{B}}(\mathbf{X})\widetilde{\mathbf{u}}^* + \boldsymbol{\eta}.
\end{equation}
The nominal MPC model evolves as
\begin{equation}
    \dot{\bar{\mathbf{X}}} = \mathbf{A\bar{X}} + \boldsymbol{\mathcal{B}}\big(\mathbf{\bar{X}}^{*}_{-1}\big)\widetilde{\mathbf{u}}^*.
\end{equation}
Consider a simplified value function for KQ-MPC
 \begin{equation}
      V(\bar{\mathbf{X}}) =  \int_{t}^{t+T_H} \! \! \big( |\mathbf{\bar{X}}(\tau)|_{\boldsymbol{\mathcal{Q}}}^2 + |\mathbf{u}(\tau)|_{\boldsymbol{\mathcal{R}}}^2 \big) d\tau.
 \end{equation}
 We define $V^*(\bar{\mathbf{X}})$ as the optimal cost-to-to such that 
 \begin{equation}
 V^*(\bar{\mathbf{X}}) = \underset{\mathbf{u}}{\min}\ V(\bar{\mathbf{X}}) 
\end{equation}
 For linear MPC schemes with controllable or stabilizable nominal model, with convex constraints and quadratic costs, asymptotic stability and recursive feasibility of Model Predictive Control (MPC) can be guaranteed without the use of terminal costs, terminal controllers, or terminal constraint sets, provided the prediction horizon $T_H$ is sufficiently long \cite{boccia2014stability}. Our nominal system model is controllable (hence stabilizable) and the MPC problem is QP with convex constraints and quadratic cost. Therefore, with sufficiently high control update, sufficiently long $T_H$, and in absence of disturbance,
 \begin{equation}
 \begin{aligned}
    \dot{V}^*(\mathbf{\bar{X}}) & = \nabla V^*(\bar{\mf{X}})\cdot(\mathbf{A}\bar{\mathbf{X}} + \boldsymbol{\mathcal{B}}\big(\bar{\mathbf{X}}^*_{-1})\widetilde{\mathbf{u}}^*\big) \\
    &\leq -\lambda_{\min}(\boldsymbol{\mathcal{Q}}) |\bar{\mathbf{X}}|^2.
    \label{eq:abd}
\end{aligned}
\end{equation}
Due to the quadratic nature of the value function, controllability of the nominal system, and under closed-loop control, it is radially bounded as
\begin{equation}
    \alpha_{1}\!\left(|\mathbf{X}|\right) \;\leq\; V^{*}(\mathbf{X}) \;\leq\; \alpha_{2}\!\left(|\mathbf{X}|\right),
\end{equation}
where $\alpha_1(\cdot)$ and $\alpha_2(\cdot)$ are class $\mathcal{K}_\infty$ functions.
Now, consider the value function along the actual system trajectories,
 \begin{equation}
 \begin{aligned}
    \dot{V}^*(\mathbf{X}) &= \nabla V^*(\mathbf{X})\cdot \dot{\mathbf{X}} \\
    & =  \nabla V^*(\mathbf{X})(\dot{\bar{\mathbf{X}}}+\dot{\mathbf{e}}_p) \\
    &=  \nabla V^*(\mathbf{X})\cdot \dot{\bar{\mathbf{X}}}+\nabla V^*(\mathbf{X})\cdot \dot{\mathbf{e}}_p\\
    &= \nabla V^*(\mathbf{\bar{X}})\cdot \dot{\bar{\mathbf{X}}} + [\nabla V^*(\mathbf{X})-\nabla V^*(\mathbf{\bar{X}})]\cdot\dot{\bar{\mathbf{X}}}\\
    & \quad \quad \quad \quad \quad \quad \quad \quad \quad \quad \quad \quad + \nabla  V^*(\mathbf{X})\cdot \dot{\mathbf{e}}_p \ . 
     \end{aligned}
    \label{eq:abd}
\end{equation}
From \eqref{eq:epdot} and \eqref{eq:ep_gr},
\begin{equation}
\begin{aligned}
    |\dot{\mathbf{e}}_p|  & \leq M_e (E_b+C_b) + L_\mathcal{B}\widetilde{u}_b \gamma_e(dt)  + \eta_b \\
    & \leq M_e \ E_b + M_e C_b +L_\mathcal{B}\widetilde{u}_b \gamma_e(dt) + \eta_b \\
    & \leq E_e + C_e,
\end{aligned}
\end{equation}
where, $E_e = M_e \ E_b + L_\mathcal{B}\widetilde{u}_b \gamma_e(dt)$ , and $C_e = M_e\ C_b+\eta_b$. Therefore, from \eqref{eq:abd}, 
\begin{equation}
    \begin{aligned}
         \dot{V}^*(\mathbf{X}) &\leq -\lambda_{\min}(\boldsymbol{\mathcal{Q}}) |\bar{\mathbf{X}}|^2 + L_V\ M_X |\mathbf{e}_p|\\
         & \quad \quad \quad \quad \quad \quad \quad  + c|\mathbf{X}|\ (E_e+C_e),
    \end{aligned}
\end{equation}
where $| \nabla V^*(\mathbf{X})-\nabla V^*(\mathbf{\bar{X}})| \leq L_V \ |\mathbf{e}_p|$ (using the local Lipschitz property of $V^*(\cdot)$ on a bounded ball $\mathbb{B}_\rho$), and $|\nabla V^*(\mathbf{X})| \leq c|\mathbf{X}|$ for any $c>0$. Also, $M_X = (A_bX_bB_b+\widetilde{u}_b)$, and $|\bar{\mathbf{X}}|^2 \geq \tfrac{1}{2}|\mathbf{X}|^2 - |\mathbf{e}_p|^2$, Therefore, with application of Young's inequalities 
\begin{equation}
    \begin{aligned}
     \dot{V}^*(\mathbf{X}) & \leq -\tfrac{\lambda_Q}{2}|\mathbf{X}|^2 +\lambda_Q(E_b+C_b)^2 +  \\
     &  \quad \quad \quad \quad L_V\ M_X(E_b+C_b) + c|\mathbf{X}|\ (E_e+C_e) \\
     & \leq - \tfrac{\lambda_Q}{2}|\mathbf{X}|^2 + (\lambda_Q+1) E_b^2 + (\lambda_Q+\lambda_Q^2)C_b^2 \\
     & \quad \quad + L_V\ M_X(E_b+C_b) + c|\mathbf{X}|\ (E_e+C_e) \\
     & \leq - \tfrac{\lambda_Q}{4}|\mathbf{X}|^2 + (\lambda_Q+1) E_b^2 + (\lambda_Q+\lambda_Q^2)C_b^2 +\\
     & \quad \quad \quad  L_V\ M_X(E_b+C_b) + \tfrac{2c^2}{\lambda_Q} E_e^2 + \tfrac{2c^2}{\lambda_Q}C_e^2,
    \end{aligned}
\end{equation}
where $\lambda_Q = \lambda_{\min}(\boldsymbol{\mathcal{Q}})$. Let,
\begin{equation}
    \begin{aligned}
        \gamma_o(dt) & = (\lambda_Q+1) E_b^2+ L_V\ M_X\ E_b + \tfrac{2c^2}{\lambda_Q} E_e^2, \\
        \sigma(\eta_b) & = \sigma(|\vect{\eta}|) =  (\lambda_Q+\lambda_Q^2)C_b^2 + L_V\ M_X\ C_b \\
        & \quad \quad \quad \quad \quad \quad \quad \quad \quad \quad \quad \quad + \tfrac{2c^2}{\lambda_Q}C_e^2, 
    \end{aligned}
\end{equation}
giving rise to the dissipation inequality
\begin{equation}
\begin{aligned}
    \dot{V}^*(\mathbf{X}) & \leq -\frac{\lambda_Q}{4}|\mathbf{X}|^2 + \sigma(\eta_b) + \gamma_o(dt)\\
    & \leq -(1-\theta)\frac{\lambda_Q}{4}|\mathbf{X}|^2 - \theta \frac{\lambda_Q}{4}|\mathbf{X}|^2 \\
    & \quad \quad \quad \quad \quad \quad \quad \quad \quad + \sigma(\eta_b) + \gamma_o(dt),\\
    & \leq -(1-\theta)\frac{\lambda_Q}{4}|\mathbf{X}|^2 \text{ when } \\
    & \quad \quad \quad \quad \quad \quad \quad \quad |\mathbf{X}| \geq 4\tfrac{\sigma(\eta_b) + \gamma_o(dt)}{\lambda_Q},
\end{aligned}
\end{equation}
where $\theta \in (0, 1)$. Therefore, as per Theorem 4.18 in \cite{khalil2002nonlinear}, the state of the Quadrotor dynamics (2) or (3) in the manuscript under KQ-LMPC scheme is \emph{uniformly ultimately bounded} by $\alpha_1^{-1}\circ\alpha_2\left(4\dfrac{\sigma(\eta_b) + \gamma_o(dt)}{\lambda_Q}\right)$, i.e., the closed-loop system is input-to-state practically stable (ISPS) with respect to the disturbance $\bs \epsilon$.

\subsection{LQR Controller Synthesis}\label{sub:lqr}
From the LTI formulation in \eqref{eq:LTI}, we consider the linearized error dynamics
\begin{equation}
    \dot{\hat{e}}(t) = \mathbf{A}\,\hat{e}(t) + \bar{\boldsymbol{\mathcal{B}}}\,\mathbf{U}(t),
\end{equation}
where $\hat{e}(t) = \mathbf{X}(t) - \mathbf{X}_r(t)$ is the state deviation from the reference $\mathbf{X}_r(t)$ (in the lifted-space). The infinite-horizon continuous-time LQR problem seeks to find the control law $\mathbf{U}(t)$ that minimizes
\begin{equation}
    J = \int_{0}^{\infty} \left( \hat{e}(t)^\top \boldsymbol{\mathcal{Q}}\,\hat{e}(t)
    + \mathbf{U}(t)^\top \boldsymbol{\mathcal{R}}\,\mathbf{U}(t) \right)\, dt,
\end{equation}
where $\boldsymbol{\mathcal{Q}} \succeq 0$ and $\boldsymbol{\mathcal{R}} \succ 0$ are symmetric weighting matrices. The optimal (lifted) control input is given by the linear state-feedback law
\begin{equation}
    \mathbf{U}^\ast(t) = -\mathbf{K}\,\hat{e}(t),
\end{equation}
where the feedback gain matrix is
\begin{equation}
    \mathbf{K} = \boldsymbol{\mathcal{R}}^{-1}\bar{\boldsymbol{\mathcal{B}}}^\top \mathbf{P}.
\end{equation}
The matrix $\mathbf{P} \succ 0$ is the unique solution of the continuous-time algebraic Riccati equation (CARE)
\begin{equation}
    \mathbf{A}^\top \mathbf{P} + \mathbf{P}\mathbf{A}
    - \mathbf{P}\bar{\boldsymbol{\mathcal{B}}}\,\boldsymbol{\mathcal{R}}^{-1}\bar{\boldsymbol{\mathcal{B}}}^\top \mathbf{P}
    + \boldsymbol{\mathcal{Q}} = 0.
\end{equation}
From \eqref{eq:least_sq}, we recover $\widetilde{\mathbf{u}}^*$ as the minimum--norm exact solution given
\begin{equation}
    \tilde{\mathbf{u}}^\star 
    = \tilde{\boldsymbol{\mathcal{B}}}^{\dagger}\mathbf{U}^\star
    = \left(\tilde{\boldsymbol{\mathcal{B}}}^\top\tilde{\boldsymbol{\mathcal{B}}}\right)^{-1}\tilde{\boldsymbol{\mathcal{B}}}^\top\mathbf{U}^\star,
\end{equation}
and apply control $\mathbf{u}^* = \mathscr{U}^{-1}(\mathbf{x},\widetilde{\mathbf{u}}^*)$ to the quadrotor. To ensure $\mathbf{u}^* \in \mathcal{U}$, we project the computed $\mathbf{u}^*$ onto $\mathcal{U}$, and apply the projected control input given as
\[
\mathbf{u}^*_{\text{clipped}} = \operatorname{proj}_{\mathcal{U}}(\mathbf{u}^*),
\]
where the projection saturates each component of $\mathbf{u}^*$ to the allowed limits.

\begin{algorithm}
\small
\caption{KQ-LMPC with LQR Fallback}\label{alg:lmpc_lqr}
\begin{algorithmic}[1]
\STATE \textbf{Input:} $\boldsymbol{\mathcal{Q}}, \boldsymbol{\mathcal{R}}, t_f, \delta, dt, \mathbf{u}_{\min},\mathbf{u}_{\max}, \mathbf{s}_{lb}, \mathbf{v}_{lb}, \boldsymbol{\omega}_{lb}, \mathbf{s}_{ub}, \mathbf{v}_{ub}, \boldsymbol{\omega}_{ub}$
\STATE \textbf{Given:} LQR feedback gain $\mathbf{K}$
\STATE $t \gets 0$
\WHILE{$0 \leq t \leq t_f$}
    \STATE Measure $\mathbf{x}(t)$ and lift to $\mathbf{X}(t) = \phi(\mathbf{x}(t))$
    \IF{$t == 0$}
        \STATE $\bar{\mathbf{X}}^*_{-1} \gets \mathbf{X}_r(l|t)$
    \ENDIF
    \STATE Attempt to solve $\mathcal{P}$ \eqref{eq:lmpc} for $\mathbf{u}^*(l|t)$
    \IF{$\mathcal{P}$ feasible}
    \STATE obtain $\bar{\mathbf{X}}^*(l|t)$
        \STATE Apply : $\mathbf{u}(\tau) \gets \mathbf{u}^*(\tau|t), \ \tau \in [t, t+\delta)$
        \STATE $t \gets t + dt$
        \STATE Store: $\bar{\mathbf{X}}^*_{-1} \gets \bar{\mathbf{X}}^*(l|t)$

    \ELSE
        \STATE $\hat{e}(t) \gets \mathbf{X}(t) - \mathbf{X}_r(t)$
        \STATE $\mathbf{U}_{\text{fb}}(t) \gets -\mathbf{K}\hat{\mathbf{e}}(t)$
        \STATE $\mathbf{u}^* \gets \tilde{\boldsymbol{\mathcal{B}}}^{\dagger}\mathbf{U}^\star$
        \STATE Apply $\mathbf{u}^*_{\text{clipped}} \gets \operatorname{proj}_{\mathcal{U}}(\mathbf{u}^*)$
        \STATE $t \gets t + dt$
        \STATE Store: $\bar{\mathbf{X}}^*_{-1} \gets 
        \mathbf{X}_r(l|t)$
    \ENDIF
\ENDWHILE
\end{algorithmic}
\end{algorithm}
\vspace{-0.6em}
\bibliographystyle{IEEEtran}
\bibliography{bibl}

\vfill

\end{document}